\documentclass[apj]{emulateapj}
\usepackage{apjfonts}

\begin{document}

\submitted{Accepted for publication in PASP}

\title{The Keck Planet Search: Detectability and the Minimum Mass and Orbital Period Distribution of Extrasolar Planets}

\author{Andrew Cumming\altaffilmark{1}, 
R.~Paul Butler\altaffilmark{2},
Geoffrey W.~Marcy\altaffilmark{3},
Steven S. Vogt\altaffilmark{4}, 
Jason T.~Wright\altaffilmark{3},
\& Debra A.~Fischer\altaffilmark{5}}

\altaffiltext{1}{Department of Physics, McGill University, 3600 rue University, Montreal QC, H3A 2T8, Canada; cumming@physics.mcgill.ca}
\altaffiltext{2}{Department of Terrestrial Magnetism, Carnegie Institute of Washington, 5241 Broad Branch Road NW, Washington, DC 20015-1305}
\altaffiltext{3}{Department of Astronomy, University of California, Berkeley, CA 94720-3411}
\altaffiltext{4}{UCO/Lick Observatory, University of California, Santa Cruz, CA 95064}
\altaffiltext{5}{Department of Physics and Astronomy, San Francisco State University, San Francisco, CA, 94132}

\begin{abstract}
We analyze 8 years of precise radial velocity measurements from the Keck Planet Search, characterizing the detection threshold, selection effects, and completeness of the survey. We first carry out a systematic search for planets, by assessing the false alarm probability associated with Keplerian orbit fits to the data. This allows us to understand the detection threshold for each star in terms of the number and time baseline of the observations, and the underlying ``noise'' from measurement errors, intrinsic stellar jitter, or additional low mass planets. We show that all planets with orbital periods $P<2000$ days, velocity amplitudes $K>20\ {\rm m\ s^{-1}}$, and eccentricities $e\lesssim 0.6$ have been announced, and we summarize the candidates at lower amplitudes and longer orbital periods. For the remaining stars, we calculate upper limits on the velocity amplitude of a companion. For orbital periods less than the duration of the observations, these are typically $10\ {\rm m\ s^{-1}}$, and increase $\propto P^2$ for longer periods. We then use the non-detections to derive completeness corrections at low amplitudes and long orbital periods, and discuss the resulting distribution of minimum mass and orbital period. We give the fraction of stars with a planet as a function of planet mass and orbital period, and extrapolate to long period orbits and low planet masses. A power law fit for planet masses $>0.3\ M_J$ and periods $<2000$ days gives a mass-period distribution $dN=C\,M^\alpha P^\beta d\ln Md\ln P$ with $\alpha=-0.31\pm 0.2$, $\beta=0.26\pm 0.1$, and the normalization constant $C$ such that 10.5\% of solar type stars have a planet with mass in the range $0.3$--$10\ M_J$ and orbital period $2$-$2000$ days. The orbital period distribution shows an increase in the planet fraction by a factor of $\approx 5$ for orbital periods $\gtrsim 300$ days. Extrapolation gives $17$--$20$\% of stars having gas giant planets within 20 AU. Finally, we constrain the occurrence rate of planets orbiting M dwarfs compared to FGK dwarfs, taking into account differences in detectability. 
\end{abstract}

\keywords{binaries: spectroscopic --- methods: statistical --- planetary systems}

\section{Introduction}

Precise Doppler velocity surveys of nearby stars have led to the detection of more than 250 extrasolar planets (e.g.~Marcy et al.~2005a; Butler et al.~2006a). They have minimum masses from $5$ Earth masses ($M_\Earth$) and up, orbital periods from close to one day up to several years, and a wide range of eccentricities. Over 25 multiple planet systems are known, with many other single planet systems showing a long term velocity trend likely indicating a second planet with long orbital period (Fischer et al.~2001). The increasing number of detections allows us to answer questions about the statistical properties of extrasolar planetary systems, such as the mass, period, and eccentricity distributions (Tabachnik \& Tremaine 2002; Butler et al.~2003; Fischer et al.~2003; Lineweaver \& Grether 2003; Jones et al.~2003; Udry, Mayor, \& Santos 2003; Gaudi, Seager, \& Mallen-Ornelas 2005; Ford \& Rasio 2006; Jones et al.~2006; Ribas \& Miralda-Escud\'e 2007), and the incidence of giant planets as a function of host star metallicity (Fischer \& Valenti 2005; Santos et al.~2005) and mass (Butler et al.~2004b; Butler et al.~2006b; Endl et al.~2006; Johnson et al.~2007).

In this paper, we focus on the frequency of planetary systems, and the distributions of mass and orbital periods. The frequency of planets is important for future astrometric and direct searches (see e.g.~Beuzit et al.~2007). The details of the mass-orbital period distribution are important because they contain information about the planet formation process (Armitage et al.~2002; Ida \& Lin 2004a, 2004b, 2005, 2008a, 2008b; Alibert et al.~2005; Rice \& Armitage 2005; Kornet \& Wolf 2006). Figure 1 shows the distribution of planet masses and orbital periods for 182 planets announced as of March 2007. Several features of the mass-period distribution have been discussed in the literature: the  ``pile-up'' at orbital periods of $\approx 3$ days (the ``hot jupiters'') (e.g.~see Gaudi et al.~2005); the paucity of massive planets ($M>1\ M_J$) in close orbits (Udry et al.~2002, 2003; Mazeh \& Zucker 2002, 2003); the deficit of planets at intermediate orbital periods of $\sim 10$ and $100$ days, giving a ``period valley'' in the orbital period distribution (Jones et al.~2003; Udry et al.~2003; Burkert \& Ida 2007); and a suggestion that the lack of lower mass planets ($M<0.75 M_J$) at orbital periods $\sim 100$--$2000$ days is significant and a real feature of the distribution (Udry et al.~2003). It has also been pointed out that the incidence and mass-period distribution of planets should depend on the mass of the host star. In particular, a much lower incidence of Jupiter mass planets is expected around M dwarfs in the core accretion scenario for planet formation (Laughlin, Bodenheimer, \& Adams 2004; Ida \& Lin 2005; Kennedy \& Kenyon 2008 although see Kornet \& Wolf 2006), and observational estimates support this picture (Butler et al.~2004b; Butler et al.~2006b; Endl et al.~2006; Johnson et al.~2007).

\begin{figure}
\plotone{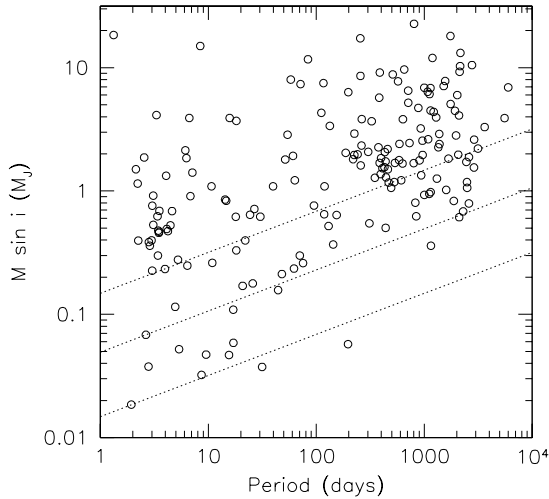}
\caption{
Minimum mass ($M\sin i$) and period ($P$) distribution of 182 extrasolar planets detected by radial velocity searches announced as of March 2007. The dotted lines show velocity amplitudes of 3, 10 and $30\ {\rm m\ s^{-1}}$ for a $1 M_\odot$ star. We take the orbital parameters from the updated Butler et al.~2006 catalog maintained at http://www.exoplanets.org.\label{fig:mp}}
\end{figure}

These interpretations are complicated by the fact that the mass-period distribution is subject to selection effects at low masses and long orbital periods. The important observational quantity is the stellar velocity amplitude induced by the planet, which for a planet of mass $M_P$ is
\begin{equation}\label{eq:K}
K={28.4\ \mathrm{m/s}\over\sqrt{1-e^2}}\ \left({M_P\sin i\over M_J}\right)\left({P\over 1\ \mathrm{yr}}\right)^{-1/3}\left({M_\star\over M_\odot}\right)^{-2/3}
\end{equation}
where $P$ is the orbital period, $e$ is the eccentricity, $M_\star$ is the mass of the star, and $i$ is the inclination of the orbit. The dotted lines in Figure \ref{fig:mp} show $K=3, 10$ and $30\ {\rm m\ s^{-1}}$ for circular orbits around a solar mass star. The detectable amplitude depends on the number and duration of the observations, and particularly on the Doppler measurement errors and other noise sources (see Cumming 2004 for a detailed discussion). 

Scatter in the measured radial velocity is expected from statistical and systematic measurement errors, and intrinsic stellar radial velocity variations or ``jitter''. The typical statistical measurement error, which we refer to in this paper as the ``Doppler error'' or ``Doppler measurement error'', is determined by the uncertainty in the mean velocity of a large number of spectral segments. The Doppler errors are typically $3$--$5\ {\rm m/s}$ in the data considered here, although Doppler errors as small as $\approx 1\ {\rm m/s}$ are now routine (Mayor et al.~2003; Butler et al.~2004a). Sources of systematic measurement errors include imperfect PSF descriptions and deconvolution algorithms, and the characterization of the charge transfer in the spectrometer CCD (see discussion in Butler et al.~2006b). Stellar jitter is thought to arise from a combination of surface convective motions, magnetic activity, and rotation (Saar \& Donahue 1997). The amount of jitter depends on stellar properties such as rotation rate and spectral type, but is typically $1$--$5\ {\rm m/s}$ for chromospherically quiet stars (Saar, Butler, \& Marcy 1998; Santos et al.~2000; Wright 2005). Additional low mass planets in a system could provide another source of radial velocity variability.

These various sources of noise determine the velocity threshold for detecting planets, and vary between observations, different stars, and different surveys. Interpretation of the mass-period distribution at low masses requires a careful analysis of these selection effects. Most work to date has taken a fixed detection threshold, such as $K=10\ {\rm m\ s^{-1}}$ (e.g.~Udry et al.~2003) or a mass cut well above the masses at which selection effects should play a role (Lineweaver \& Grether 2003). A few detailed calculations of detection thresholds have been carried out. In Cumming, Marcy, \& Butler (1999), we presented an analysis of 11 years of Doppler measurements of 76 stars as part of the Lick planet search. However, the conclusions regarding the mass-period distribution were limited because only 6 planets were then known. Endl et al.~(2002) present a statistical analysis of the 37 star sample observed by the ESO Coud\'e Echelle spectrometer. Wittenmyer et al.~(2006) present limits on companion mass for 31 stars observed at McDonald Observatory. The largest study so far is that of Naef et al.~(2005), who derive detection thresholds for 330 stars from the ELODIE Planet Search and estimate planet occurrence rates. 

In this paper, we analyse 8 years of radial velocity measurements from the Keck survey, consisting of data taken from the beginning of the survey in 1996 to the time of the HIRES spectrometer upgrade in 2004 August. The number of stars ($585$) and planets ($48$) included in the sample offer an order of magnitude improvement over our previous Cumming et al.~(1999) analysis, and therefore the best opportunity to date to determine the occurrence rate of planets and their mass-period distribution. An outline of the paper is as follows. In \S 2, we describe a technique for identifying planets in radial velocity data, discuss the detection thresholds for the survey, and calculate limits on the mass and period of planets orbiting stars that do not have a significant detection. In \S 3, we use these limits to correct the mass and period distributions for incompleteness, and then characterize the occurrence rate of planets and the mass-period distribution. We summarize and conclude in \S 4.

\section{Search for Planets}

\subsection{Observations}

\begin{figure}
\epsscale{1.0}\plotone{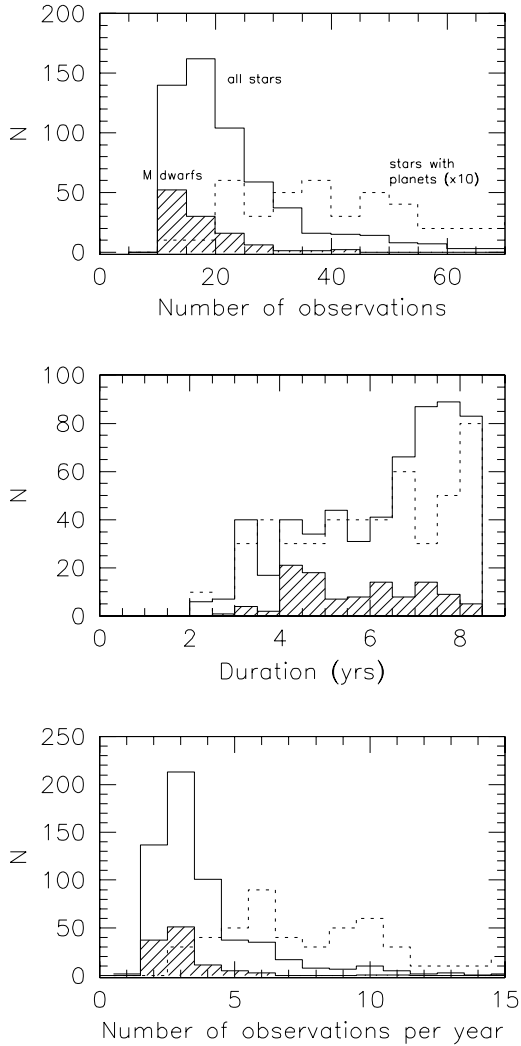}
\caption{The number, duration, and average rate of observations, for the 585 stars in the sample (all of which have 10 or more observations over a period of two years or more). The shaded histograms are for the subset of stars with $M_\star<0.5M_\sun$. The dotted histograms are for the subset of 48 stars with an announced planet, and have been multiplied by a factor of 10 for clarity.
\label{fig:obs1}}
\end{figure}

The Keck Planet Search program has been in operation since 1996 July, using the HIRES echelle spectrometer on the Keck I telescope (Vogt et al.~1994; Vogt et al.~2000). The data we analyze here were taken from the beginning of the survey up to 2004 August when the HIRES spectrometer was upgraded. They consist of radial velocity measurements of 585 F, G, K, and M stars (the fractions in these spectral classes are 7, 49, 27, and 16\% respectively). Note that the M dwarf sample covers spectral types M5 and earlier; the F stars are of spectral type F5 and later. Selection of the target stars is described in Wright et al.~(2004) and Marcy et al.~(2005b). They lie close to the main sequence and are chromospherically quiet. They have $B-V>0.55$, declination $>-35^\circ$, and have no stellar companion within 2". To ensure enough data points for an adequate Keplerian fit to the data, we further select only those stars with at least 10 observations over a period of two years or more. This excludes data for an additional 360 stars that were added in the two years prior to 2004 August. Figure \ref{fig:obs1} is a summary of the number, time baseline or duration, and mean rate of observations. Typical values are $10$--$30$ observations in total over a duration of $6$--$8$ years, with 3 observations per year. The target list of stars has changed over the years of the survey, with stars being dropped and added (see Marcy et al.~2005b for a discussion of the evolution of the target list), resulting in the spread in durations shown in Figure \ref{fig:obs1}.

\begin{figure}
\epsscale{1}\plotone{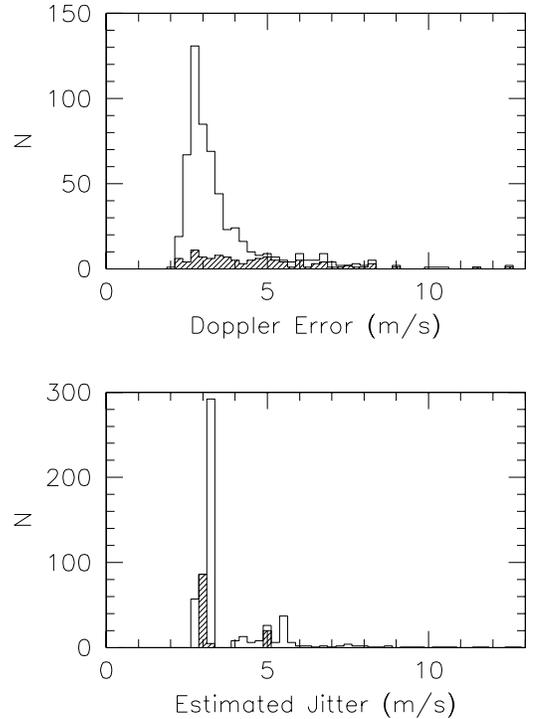}
\caption{The mean Doppler velocity measurement error, and estimated jitter, for the stars in the sample. The jitter estimates are taken from Wright (2005) and include both intrinsic stellar jitter and systematic measurement errors. The shaded histograms are for the subset of stars with $M_\star<0.5M_\sun$.\label{fig:obs2}}
\end{figure}

Figure \ref{fig:obs2} is a summary of the statistical Doppler measurement errors and the estimated jitter from Wright (2005). The Doppler error shown is the mean Doppler error for all observations for each star. These are statistical errors determined by the weighted uncertainty in the mean velocity of 400 spectral segments, each 2\AA\ long. Most stars have a mean Doppler measurement error of $\approx 3\ \mathrm{m/s}$ (smaller errors of $\approx 1\ {\rm m\ s^{-1}}$ have been achieved at Keck following the 2004 HIRES upgrade, but these data are not included in this analysis). The stars in the sample have been selected to be chromospherically inactive, but still show stellar jitter at the few ${\rm m\ s^{-1}}$ level (Saar et al.~1998).  We adopt the jitter estimates described in Marcy et al.~(2005b) and Wright (2005), in which the level of jitter is calibrated in terms of stellar properties, in particular $B-V$, $M_V$, and $R_{HK}$. These values include contributions from both intrinsic stellar jitter and systematic measurement errors. The typical jitter is $\sigma_{\rm jitter}\approx 3\ \mathrm {m/s}$ for a chromospherically quiet star, with a tail of larger values for more active stars. This is comparable to the Doppler measurement errors, indicating that the velocity measurements have reached a precision for which stellar jitter and systematic errors are beginning to dominate the uncertainty.

In Figures \ref{fig:obs1} and \ref{fig:obs2} the dotted histograms summarize the observations of the 110 stars with stellar masses $M_\star<0.5\ M_\sun$, which are almost entirely M dwarfs. In \S \ref{sec:mp}, we analyze the mass-orbital period distribution separately for stars with $M_\star>0.5 M_\sun$ and $M_\star<0.5 M_\sun$, since the planet occurrence rate for M dwarfs appears to be smaller than that for FGK stars. Figures \ref{fig:obs1} and \ref{fig:obs2} show that in general the Doppler errors for the M dwarfs are larger than for the FGK stars, mostly due to their relative faintness. The dashed histograms in Figures \ref{fig:obs1} are for the subset of stars with an announced planet (see \S 2.4). Once a candidate planetary signature is detected in the data, we increase the priority of that star in our observing schedule, resulting in a greater number of observations for those stars with announced planets.

\begin{figure}
\epsscale{1.0}\plotone{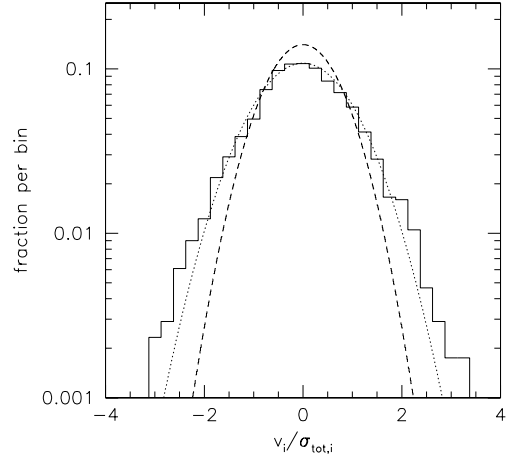}
\caption{The distribution of measured velocity $v_i$ normalized by the estimated variability $\sigma_{{\rm tot},i}$, the quadrature sum of Doppler measurement errors and estimated stellar jitter. The velocities used here are for a subset of quiet stars, after subtracting the mean for each data set. The dashed histogram shows a unit Gaussian. 
We compare the observed distribution (solid curve) with a Gaussian with errors increased by 30\% (dotted curve).\label{fig:normres}}
\end{figure}

We include the Doppler errors and jitter by adding them in quadrature to find a total estimated error for each data point $i$, $\sigma_{{\rm tot},i}^2=\sigma_i^2+\sigma_{\rm jitter}^2$. Figure \ref{fig:normres} shows the distribution of the residuals after subtracting the mean velocity for a set of 386 ``quiet'' stars which after a preliminary analysis show no excess variability, long term trend, or evidence for a periodicity. We plot a histogram of the ratio $v_i/\sigma_{{\rm tot}, i}$ for all 3436 velocities for these stars, compared with a normal distribution. The width of the observed distribution is $\approx 30$\% greater than a unit variance Gaussian, suggesting that the estimated variability is underpredicted by this factor. To allow for this, we have multiplied each $\sigma_{{\rm tot},i}$ by a factor of 1.3 in the analysis that follows. The dotted histogram shows a normal distribution with $\sigma=1.3$ for comparison with the observed distribution. The Gaussian distribution underpredicts the tails of the distribution somewhat, but otherwise agrees quite well. The magnitude of this correction has only a small effect on the results of this paper, because when assessing the significance of a Keplerian orbit fit, we calculate ratios of chi-squared (for example, the ratio of chi-squareds with and without the Keplerian orbit included in the model) and so the role of the errors $\sigma_{{\rm tot},i}$ is to set the relative weighting of different data points (see Cumming et al.~1999 for a discussion).

\subsection{Long Term Trends and Excess Variability}

The first indication of a companion is often excess velocity variability above the level expected from measurement errors and stellar jitter. To assess this, we fit a straight line to each data set, and compare the observed scatter about the straight line to the predicted value. The data for each star are a set of measured velocities $\left\{v_i\right\}$, observation times $\left\{t_i\right\}$, and estimated errors $\left\{\sigma_{{\rm tot},i}\right\}$. The estimated error $\sigma_{{\rm tot},i}$ is the quadrature sum of the Doppler error and stellar jitter, as described in \S 2.1. We fit either a constant ($f_i=a$) or a straight line ($f_i=a+bt_i$) to the data. To test whether including a slope significantly improves the chi-squared of the fit,
\begin{equation}
\chi^2=\sum_i \ {(v_i-f_i)^2\over \sigma_{{\rm tot},i}^2},
\end{equation}
we use an F-test (Bevington \& Robinson 1992). The appropriate F-statistic is
\begin{equation}
F_{(N-2),1}={(N-2)}\left({\chi^2_{\rm constant}-\chi^2_{\rm slope}\over\chi^2_{\rm slope}}\right)
\end{equation}
(Bevington \& Robinson 1992), where $\chi^2_{\rm constant}$ is the $\chi^2$ for the best-fitting constant, and $\chi^2_{\rm slope}$ is the $\chi^2$ for the best-fitting straight line. We determine the probability that the observed value of $F_{(N-2),1}$ is drawn from the corresponding $F$-distribution. If this probability is smaller than $0.1$\%, we conclude that the slope is significant. We make the choice of $0.1$\% so that we expect no false detections in our sample of 585 stars. We find that 95 stars (16\% of the total) have a significant slope. This fraction is consistent with the 20 out of 76 stars in the Lick sample, or 26\%, that showed a long-term trend at the 0.1\% significance level (Cumming et al.~1999).

Having decided whether a constant or a straight line best describes the long term behavior, we then test whether the residuals to the fit are consistent with the expected variability. We calculate the probability that $\chi^2_{\rm constant}$ or $\chi^2_{\rm slope}$ is drawn from a chi-squared distribution with $N-1$ or $N-2$ degrees of freedom respectively (Hoel, Port, \& Stone 1971). We again choose a 0.1\% threshold, so if this probability is smaller than 0.1\% we infer that there is excess variability in the data. We find that of the 585 stars, 131 show significant variability at the 0.1\% level, or 23\% of the total. Of these 131 stars, 34 (26\%) also show a significant slope (i.e.~6\% of the 585 stars show both significant variability and a significant slope). This is similar to the Lick survey results, where we found 17 out of 76 stars, or 22\%, with excess variability (Cumming et al.~1999; see also Nidever et al.~2002 who found that 107 out of 889 stars showed velocity variations of more than $100\ {\rm m\ s^{-1}}$ over 4 years).

\subsection{Keplerian Fitting}

To search for the signature of an orbiting planet, we fit Keplerian orbits to the radial velocities, and assess the significance of the fit (Cumming 2004, hereafter C04; Marcy et al.~2005b; O'Toole et al.~2007). The non-linear least-squares fit of a Keplerian orbit requires a good initial guess for the best-fitting parameters since there are many local minima in the $\chi^2$ space. Our approach is to first limit the fit to circular orbits, and use the best-fitting circular orbit parameters as a starting point for a full Keplerian fit. 

It is important to note here that we search only for a single planet. We do not consider multiple planet systems in this paper, therefore our results for the mass and orbital period distributions apply only to the planet with the highest Doppler amplitude in a given system. To be consistent in this approach, we do not include any long term trends detected in \S 2.2 in the planet fits, and compare only to a constant velocity when assessing the significance of a given Keplerian fit. A star with a significant long-term trend will therefore be flagged as having a significant Keplerian fit with orbital period much longer than the duration of the observations. The multiplicity of planets as a function of their orbital periods is an important question that we leave to future work.

To find the best-fitting circular orbit, we calculate the Lomb-Scargle (LS) periodogram (Lomb 1976; Scargle 1982) for each data set. This involves fitting a sinusoid plus constant\footnote{Note that we extend the original LS periodogram by allowing the mean to ``float'' at each frequency,  following Walker et al.~(1995), Nelson \& Angel (1998), and Cumming et al.~(1999), rather than subtracting the mean of the data prior to the fit.} to the data for a range of trial orbital periods $P$. For each period, the goodness of fit is indicated by the amount by which including a sinusoid in the fit lowers $\chi^2$ compared to a model in which the velocity is constant in time. This is measured by the periodogram power
\begin{equation}
z(\omega)={(N-3)\over 2}\left({\chi^2_{\rm constant}-\chi^2_{\rm circ}(\omega)\over \chi^2_{\rm circ}(\omega_0)}\right),
\end{equation}
where $\chi^2_{\rm constant}=\sum_i(v_i-\langle v\rangle)^2/\sigma_{{\rm tot}, i}^2$ is the $\chi^2$ of a fit of a constant to the data, $\chi^2_{\rm circ}$ is  the $\chi^2$ of the circular orbit fit, $\langle v\rangle$ is the mean velocity, $\omega=2\pi/P$ is the trial orbital frequency, and $\omega_0$ the best-fitting frequency. The number of degrees of freedom in the sinusoid fit is $N-3$. A large power $z$ indicates that including a sinusoid significantly decreased the $\chi^2$ of the fit. We consider trial orbital periods between 1 day and 30 years.

We next choose two best fitting solutions as starting points for a full non-linear Keplerian fit. The two sinusoids are chosen so that they are well-separated in frequency. We then use a Levenberg-Marquardt algorithm (Press et al.~1992) to search for the minimum $\chi^2$, starting with a Keplerian orbit with the same period and amplitude as the sinusoid fits, and trying several different choices for the time of periastron, eccentricity, and longitude of pericenter. Having obtained the minimum $\chi^2$ from the Keplerian fit, we define a power $z_e$ to measure the goodness of fit analogous to the LS periodogram power for circular orbits,
\begin{equation}\label{eq:ze}
z_e(\omega)={(N-5)\over 4}\left({\chi^2_{\rm constant}-\chi^2_{\rm Kep}(\omega)\over \chi^2_{\rm Kep}(\omega_0)}\right)
\end{equation}
(C04), where $\chi^2_{\rm Kep}$ is the $\chi^2$ of a fit of a Keplerian orbit to the data (with $N-5$ degrees of freedom).

The significance of the fit depends on how often a power as large as the observed power $z_0$ would arise purely due to noise alone (C04; Marcy et al.~2005b). For a single frequency search, the distribution of powers due to noise alone can be written down analytically for Gaussian noise, it is
\begin{equation}\label{eq:keplerprob}
{\rm Prob}(z>z_0)=\left(1+{(\nu+2)\over 2}{4z_0\over \nu}\right)\left(1+{4z_0\over\nu}\right)^{-(\nu+2)/2}
\end{equation}
(C04), where $\nu=N-5$. However, the total false alarm probability depends on how many independent frequencies are searched. For a search of many frequencies, each independent frequency must be counted as an individual trial. The false alarm probability (FAP) is 
\begin{equation}\label{eq:FA}
F=1-\left[1-{\rm Prob}(z>z_0)\right]^{N_f}\approx N_f\ {\rm Prob}(z>z_0),
\end{equation}
where $N_f$ is the number of independent frequencies, and in the second step we assume $F$ is small. For small $F$, the FAP is simply the single frequency FAP multiplied by the number of frequencies.

An estimate of the number of independent frequencies is $N_f\approx T\Delta f$, where $\Delta f=f_2-f_1$ is the frequency range searched and $T$ is the duration of the data set (C04). For evenly-sampled data, the number of independent frequencies is $N/2$, ranging from $1/T$ to the Nyquist frequency $f_{Ny}=N/2T$. For unevenly-sampled data, Horne \& Baliunas (1986) found that $N_f\sim N$ for a search up to the Nyquist frequency (see also Press et al.~1992). This agrees with our simple estimate since $N_f\approx f_2T\approx N/2$. However, uneven sampling allows frequencies much higher than the Nyquist frequency to be searched (see discussion in Scargle 1982). In general, $N_f\gg N$, by a factor of $\approx f_2/f_{Ny}$. For example, a set of 30 observations over 7 years has $f_{Ny}\approx 1/(6\ {\rm months})$. A search for periods as short as 2 days therefore has $N_f\approx N(6\ {\rm months}/2\ {\rm days})\approx 90N\approx 2700$. Therefore to detect a signal with a FAP of $10^{-3}$, the periodogram power for that signal must be large enough, or the Keplerian fit good enough, that the single-trial false alarm probability is $\sim 10^{-6}$.

The estimate for $N_f$ and equations (\ref{eq:keplerprob}) and (\ref{eq:FA}) allow an analytic calculation of the false alarm probability (see Fig.~2 of C04). To check this analytic estimate, we determine the FAP and $N_f$ numerically using Monte Carlo simulations. The disadvantage of calculating the FAP in this way is that it is computationally intensive for Keplerian fits. This is the reason why we consider only the two best-fitting sinusoid models as starting points for the Keplerian fit. We find that the analytic estimate of the FAP agrees well with the FAP determined by the Monte Carlo simulations. Our method is to generate a large number $N_{\rm trials}$ of data sets consisting of noise only, using the same observation times as the data, and calculate the maximum power for each of them. The fraction of trials for which the maximum power exceeds the observed value then gives the false alarm probability. In addition, by fitting equation (\ref{eq:FA}) to the numerical results, we can determine $N_f$. This allows a determination of the FAP even when it is much smaller than $1/N_{\rm trials}$ (C04). We generate the noise in two ways, which give similar results for the FAP (see Cumming et al.~1999): (i) by selecting from a Gaussian distribution with standard deviation given by the observed error $\sigma_{{\rm tot},i}$ for each observation time, or  (ii) by selecting with replacement from the observed velocities (after first subtracting the mean). The second approach (a ``bootstrapping'' method, see Press et al.~1992) has the advantage that it avoids the assumption of Gaussian noise; instead the actual velocity values are used as an estimate of the noise distribution. It is similar to the  ``velocity scrambling''  method of Marcy et al.~(2005b) for determining the false alarm probability for Keplerian fits, with the main difference being that we select with replacement from the observed velocities rather than randomizing the order of the observations.

\subsection{Significant Detections}

\begin{figure}
\epsscale{1.0}\plotone{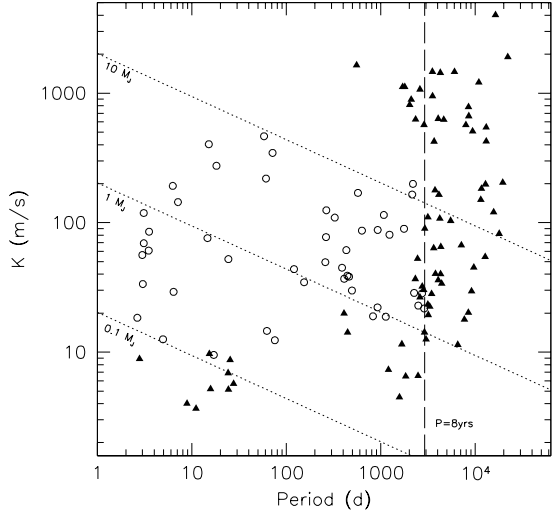}
\caption{The results of the search for significant Keplerian fits. The solid triangles (76 points) and open circles (48 points) show periodicities with FAP$<10^{-3}$. The open circles indicate stars that have an announced planet. The vertical dashed line shows $P=8$ years, corresponding to the duration of the survey. Dotted lines show the velocity amplitude corresponding to $M\sin i=0.1,1,$ and $10\ M_J$ for a solar mass star.\label{fig:keck2}}
\end{figure}

Figure \ref{fig:keck2} shows the results of the search for significant Keplerian fits. We plot the best-fit amplitude and period for stars with FAP$<0.1$\%, divided into two categories. The open circles are for stars with an announced planet (i.e. a published orbital solution) as of 2005 May; the solid triangles are stars with a significant Keplerian fit, but not confirmed as a planet\footnote{We choose 2005 May as the cutoff for publication of an orbital solution so that the announcement of the planet is based only on the data considered here, which is taken before 2004 August. In fact, 7 of our 78 candidates have since been announced based on additional data collected since 2004 August. They are four of the five planets announced by Butler et al.~(2006a), the planet orbiting the M dwarf GJ~849 (Butler et al.~2006b), and the two long period companions with incomplete orbits described by Wright et al.~(2007).}. We will refer to these significant detections that are not announced as planets as  ``candidate'' planets. There are 48 announced planets detected in our search (8.2\% of the total number of stars), and 76 candidates (13\% of the total). All but five of the announced planets in our sample of stars were detected by our algorithm (the 5 planets that were not detected orbit stars that were added to the survey only recently to confirm detections by other groups, see discussion below).

Of the 76 candidates, 27 have large velocity amplitudes $K\gtrsim 200\ {\rm m\ s^{-1}}$ corresponding to companion masses $\gtrsim 20\ {\rm M_J}$. The remaining 49 candidates, which have $M_P\sin i<15\ M_J$, fall into two groups: they are either at long orbital periods ($P\gtrsim 2000$ days), or low amplitudes ($K\lesssim 10$--$20\ {\rm m\ s^{-1}}$) compared with the announced planets. That these are the detections that have not yet been announced makes sense since (i) it is difficult to constrain orbital parameters for a partial orbit, and so we generally wait for completion of at least one full orbit before announcing a planet, and (ii) for low amplitude signals, it becomes difficult to disentangle possible false signals from stellar photospheric jitter or from systematic variations in the measured velocities.

\begin{figure}
\epsscale{1.0}\plotone{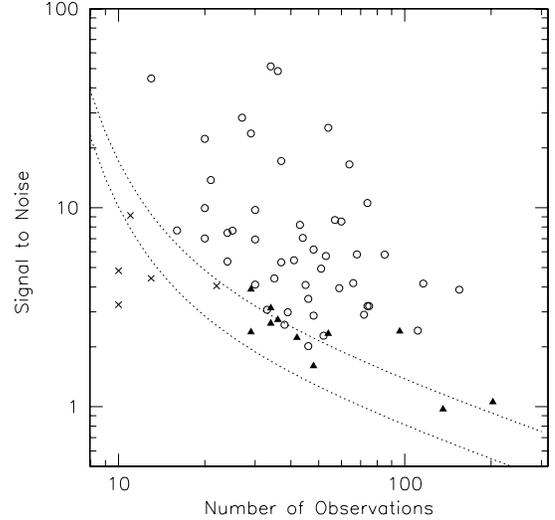}
\caption{Parameter space of signal to noise ratio $K/s$ against number of observations for the significant detections. The ``noise'' is the standard deviation of the residuals to the best-fit Keplerian orbit. We plot only those points with periods $<1000$ days, and planet mass $<15$ Jupiter masses, for which signal to noise is expected to be the limiting factor.  The open circles indicate stars that have an announced planet. The solid triangles show candidate planets which have FAP$<10^{-3}$. The crosses show the five planets that were announced by other groups, but not flagged as significant in our search, which includes only a small number of observations for these stars. The dotted curves show analytic results for the signal to noise needed for a detection probability of 50\% (lower curve) and 99\% (upper curve) (see eq.~[8]; we assume $N_f/F=10^6$).
\label{fig:sn}}
\end{figure}

In the first case (long orbital periods), it is simple to understand the detection limit, which is set by the time baseline of the observations. The vertical dashed line in Figure \ref{fig:keck2} shows an orbital period of $8$ years (the duration of the longest data set considered here), and nicely divides the announced and candidate planets. In the second case (low amplitudes), an interesting question to consider is how the threshold for announcing a planet relates to the statistical threshold for detecting a Keplerian orbit. The statistical detection threshold depends on both the number of observations and signal amplitude. For circular orbits, an analytic expression for the signal to noise required to detect a signal with 50\% probability is given by
\begin{equation}\label{eq:sn1}
{K\over\sqrt{2}s}=\left[\left({N_f\over
F}\right)^{2/\nu}-1\right]^{1/2}
\end{equation}
(Cumming et al.~2003; C04), where $F$ is the false alarm probability associated with the detection threshold, $N_f$ is the number of independent frequencies, $s$ is the noise level, and $\nu=N-3$. The number of independent frequencies $N_f$ is set by the number of observations $N$ and the duration of the observations $T$, as we discussed in \S 2.3. Figure \ref{fig:sn} compares the signal to noise ratio for each detection with this analytic result. For this comparison, we define the noise $s$ to be the rms amplitude of the residuals to the best fit orbit, and the signal to noise ratio as $K/s$. We show only those detections with best fitting period $P<1000$ days and mass $M_p\sin i<15 M_J$, for which signal to noise is expected to be the main limiting factor. The dotted curves show the analytic result for a detection probability of 50\% and 99\%.

The candidate detections mostly fall near the detection curves in Figure \ref{fig:sn}, whereas the announced planets lie above the curves. The five crosses represent planets that were detected by other groups. Their host stars, HD~8574, HD~74156, HD~82943, HD~130322, and HD~169830, were added to the Keck survey to confirm these detections, but do not yet have enough observations for a detection. They all lie below the dotted curves. 

Inspection of Figure \ref{fig:sn} shows that the detection threshold is determined by statistics when the signal to noise ratio is larger than $2$--$3$. For lower amplitude signals, the statistical significance is no longer enough, since as we mentioned there is the danger that the observed variation is in fact from stellar jitter or systematic effects. Figure \ref{fig:sn} shows that the effective detection threshold is then at a signal to noise of $\approx 2$. To detect a planet with a lower amplitude than this requires significantly more work to rule out false signals.

Comparison with the published orbits shows that our automated technique reproduces the fitted orbital parameters well, except for 2 of the 48 announced planets. For HD~50499, we find an orbital period of $2\times 10^4$ days rather than 3000 days. We include a slope in the fit for this star, which reproduces the published orbital parameters. We find a period of 15 rather than 111 days for the highly eccentric planet around HD~80606 (which has $e=0.93$; Naef et al.~2001). These examples illustrate the weakness of the Lomb-Scargle periodogram at providing a good initial guess for the Keplerian fit, in particular for eccentric orbits, and emphasises the importance of trying many different starting periods. There are also strong aliasing or spectral leakage effects at 1 day, and so when fitting Keplerian orbits, we force the orbital period to be longer than 1.2 days rather than the 1 day limit of our periodogram search. We might also expect the search to fail for multiple planet systems; however, in all cases of announced multiple planet systems, we find a significiant single planet fit, usually for the planet with the largest velocity amplitude. This is the case even when the periods are close to or are harmonically related. For example, HD~128311 has two planets in a 2:1 resonance (Vogt et al.~2005). We detect the longer period planet in our search.

\begin{figure}
\epsscale{1}\plotone{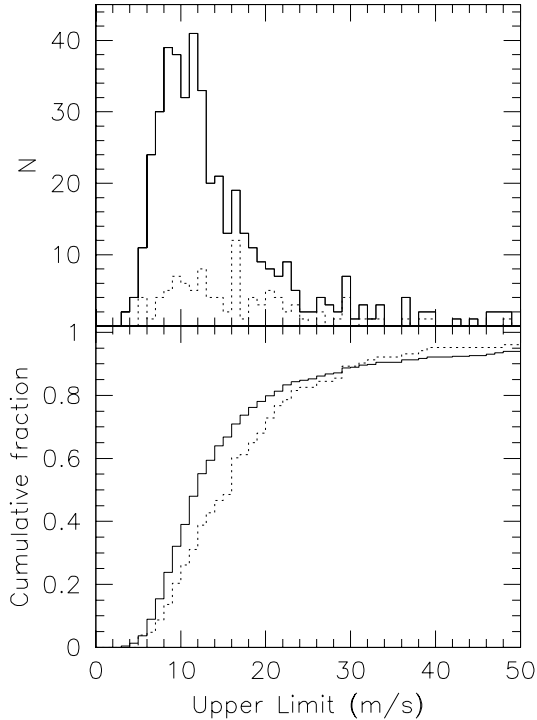}
\caption{Upper panel: histogram of the 99\% upper limit on velocity amplitude $K$ of circular orbits for 461 stars without significant detections. The upper limit is averaged over orbital periods smaller than the duration of the observations ($P<T$). Lower panel: the fraction of stars whose upper limit is less than a given value of $K$. In each panel, the solid histogram is for the entire sample of stars. The dotted histogram is for the 103 stars with $M_\star<0.5 M_\sun$. \label{fig:up}}
\end{figure}

\subsection{Upper Limits}

We next calculate upper limits on the radial velocity amplitude of planets for those stars without a significant detection. To reduce the computational time and because we are not considering the eccentricity distribution in detail in this paper, we place upper limits on the amplitude of circular orbits only. Endl et al.~(2002) and C04 showed that the detectability of a planet in an eccentric orbit is only slightly affected by eccentricity for $e\lesssim 0.5$, but can be substantially affected for larger eccentricities if the phase coverage is inadequate (e.g.~see Fig.~6 of C04). About 14\% of the known planet population has $e\geq 0.5$. In our sample, 6 stars out of 48 announced planets have $e>0.5$ (13\%), and 3 have $e>0.6$ (6\%). However, the true fraction with eccentricities greater than 0.5 may be larger because of the selection effects acting against highly eccentric orbits. Of the 76 candidate periodicities, 30 have $e>0.5$, and 11 have $e>0.6$. The larger fraction of eccentric orbits for these candidates than for the announced planets is likely due to the long orbital periods of many of the candidates for which the orbital eccentricity is not well-constrained. We leave a discussion of the eccentricity distribution to a future paper, and here assume circular orbits.

For all stars with FAP$>10^{-3}$, we calculate upper limits as described in Cumming et al.~(1999), utilizing the LS periodogram for sinusoid fits. At a given orbital period, we generate fake data sets of a sinusoid plus Gaussian noise. We assume that the amplitude of the noise is equal to the rms of the residuals to the best fitting sinusoid for the actual data. We then find the sinusoid amplitude that gives a LS periodogram power larger than the observed value in 99\% of trials. We calculate the upper limit on $K$ as a function of orbital period. However, the upper limit is insensitive to period for $P<T$ because the uneven sampling gives good phase coverage for most periods (Scargle 1982; Cumming et al.~1999). For $P>T$, the 99\% upper limit scales close to $K\propto P^2$ (for periods $T\lesssim P\lesssim 100\pi T/8\approx 300$ years; see C04).

Figure \ref{fig:up} is a histogram of the mean upper limit on $K$ for orbital periods shorter than the duration of the observations ($P<T$)\footnote{We average the upper limits over period here only to summarize the results of our calculation. In the next section, where we correct for incompleteness, we do not average over period but instead use the upper limits calculated as a function of period.}. Most stars have upper limits to $K$ of between $10$ and $15\ {\rm m\ s^{-1}}$. Taking a typical value for the quadrature sum of jitter and Doppler errors to be $\sigma\approx 3\sqrt{2}\ {\rm m\ s^{-1}}$ (since both jitter and Doppler errors are typically $3\ {\rm m\ s^{-1}}$), these upper limits correspond to signal to noise ratios $K/\sigma$ of between 2 and 3. This compares well with the analytic formula in equation (\ref{eq:sn1}) which gives $K\approx 2\sigma$ for $N=20$ and $N_f=1000$. Notice that $\approx 10$\% of stars have upper limits $>30\ {\rm m\ s^{-1}}$. This is due to either large Doppler measurement errors for these stars, or a number of observations that is too small to give a good limit. The dotted histogram shows the upper limits for the M dwarfs only ($M_\star<0.5\ M_\sun$). The upper limits are on average higher for these low mass stars. Radial velocity measurements are more difficult for M dwarfs because they are much fainter than solar mass stars.

\begin{figure}
\epsscale{1.0}\plotone{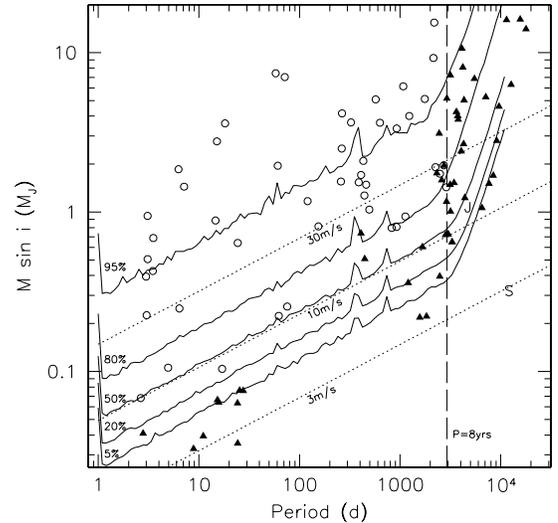}
\caption{Significant periodicities (FAP$<10^{-3}$; open circles and solid triangles) in the mass-period plane. For a given period, the solid curves show the mass that can be ruled out at the 99\% level from 5\%, 20\%, 50\%, 80\%, and 95\% of stars. Note that whereas the search for planets involves full Keplerian fits to the data, the upper limits are for circular orbits only (applicable to Keplerian orbits with $e\lesssim 0.5$, see discussion in \S 2.5). The dotted lines show velocity amplitudes $K=3$, $10$, and $30\ {\rm m\ s^{-1}}$ for a $1\ M_\odot$ star. The vertical dashed line shows $P=8$ years, corresponding to the duration of the survey. J and S label the locations of Jupiter and Saturn in this plot.\label{fig:keckm}}
\end{figure}

\subsection{Summary of Results}

Figure \ref{fig:keckm} is a summary of the analysis in this section in the mass-orbital period plane. To convert between velocity amplitude $K$ and $M_P\sin i$, we require the stellar mass (eq.~[1] gives $M_P\sin i\propto K M_\star^{2/3}$). For the stars with significant Keplerian fits, either announced or candidate planets, we use the latest mass estimates given in the Takeda et al.~(2007) and Valenti \& Fisher (2005) catalogs (except for 7 of the candidate stars that are not listed in Takeda et al.~2007 or Valenti \& Fischer 2005). For convenience, approximate stellar masses of the remaining stars are determined using the B$-$V stellar mass relation given in Allen (2000)\footnote{This relation underestimates the stellar mass for some stars, typically by $30$\% but sometimes as much as a factor of 2, for those stars which are metal rich. Therefore we expect the blue curves showing upper limits on $M_P\sin i$ as a function of period averaged over all stars in Figs.~8 and 9 to be uncertain at the $\sim 20$\% level because of this approximation. However, note that this uncertainty does not affect the completeness corrections and mass and period distributions calculated in \S 3 because they depend on the velocity upper limit directly. The only stellar mass information needed in \S 3 is for the announced and candidate planets, for which we use the accurate Takeda et al.~(2007) or Valenti \& Fischer (2005) masses.}.

We show the detections with FAP $<10^{-3}$ in Figure \ref{fig:keckm} as solid triangles and open circles, dividing them into candidates and announced planets respectively. As expected from the discussion in \S 2.4, the candidate periodicities (solid triangles) are mainly concentrated at $K\lesssim 10\ {\rm m\ s^{-1}}$ or at $P>1000$ days. The solid curves in Figure \ref{fig:keckm} summarize the upper limits as a function of period. For a given period, they show the mass which can be excluded at the 99\% level from 5\%, 20\%, 50\%, 80\%, and 95\% of stars. The aliasing effects are not strong: the upper limits vary smoothly with period because of the uneven time sampling which gives good phase coverage at most orbital periods (e.g.~Scargle 1982).  However, there is a slight reduction in sensitivity at periods close to 1 year, 2 years, and 1 day. The curves turn upwards and scale as $P^2$ for periods beyond $\approx 3000$ days ($\approx 8$ years), close to the duration of the observations.

The results shown in Figure \ref{fig:keckm} allow us to draw a number of conclusions. First, there are many candidate gas giants in orbital periods of 5--20 years, similar to our Solar System. Figure \ref{fig:keckm} shows that the main limitation at the moment for detection of an analog of our Solar System is the duration of the survey, rather than the sensitivity. We suspect that some of these long period candidates will turn out to be more massive that the $M\sin i$ indicated in Figure \ref{fig:keckm}, since only a partial orbit has been observed, leaving the fitted mass uncertain. There are several candidates with $K<10\ {\rm m\ s^{-1}}$. For these candidates, further observations are needed to rule out stellar jitter as the cause of the observed periodic variability. The upper limit curves continue smoothly to periods smaller than 3 days, and are not noticeably affected until they get close to 1 day. This implies that the abrupt drop in the number of planets at $P_{\rm orb}\lesssim 3$ days is a real feature. In the ``period valley'' between 10--100 days (Jones et al.~2003; Udry et al.~2003), the detectability is good, with upper limits of $\approx 0.1$--$0.2 M_J$ for 50\% of stars. However, for periods $>100$ days, the upper limits are larger, $0.3$--$0.6 M_J$. Therefore the reported lack of objects with $M<0.75 M_J$ at $P>100$ days by Udry et al.~(2003) is clearly dependent on understanding the selection effects in this region. We address this in the next section by using these upper limits to correct the observed period and mass distributions for incompleteness.

\begin{figure}
\epsscale{1.0}\plotone{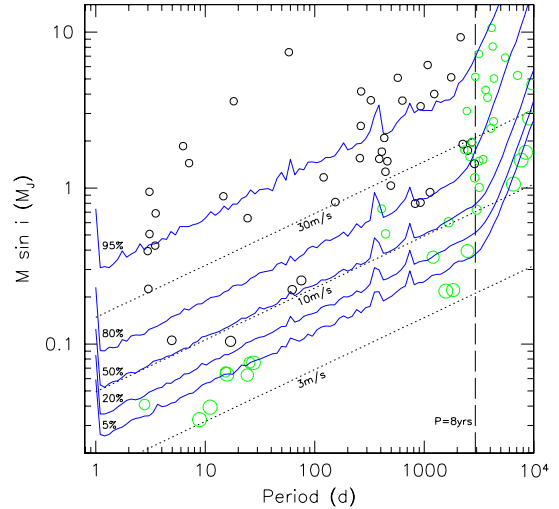}
\caption{Detections for F, G, and K stars corrected for completeness of the survey. We show the announced planets (those published by 2005 May; black circles) and candidates (significant detections not corresponding to a published planet; green circles), with the area of the circle indicating the size of the completeness correction $N_i$ for each point. The vertical dashed line indicates a period of 8 years, roughly equal to the duration of the survey. The dotted lines show velocity amplitudes of $3,10$ and $30\ {\rm m\ s^{-1}}$ for a $1\,M_\odot$ star. The blue curves summarize the upper limits as in Fig.~\ref{fig:keckm}.\label{fig:avnim}}
\end{figure}

\section{The mass-period distribution}\label{sec:mp}

In this section, we describe a method for correcting for incompleteness by taking into account the non-detections, and discuss the resulting distribution of minimum masses and orbital periods. For conciseness we  will refer to the minimum mass $M_P\sin i$ as ``mass'' throughout this section, although it should be noted that we do not include the distribution of inclination angles $i$ which is needed to determine the true mass from the measured minimum mass (Jorissen, Mayor, \& Udry 2001; Zucker \& Mazeh 2001). This is reasonable for a large statistical sample. The average value of the ratio of minimum mass to true mass is $\langle\sin i\rangle=\pi/4=0.785$, a small correction for this analysis, and we also note that power law scalings are not affected by the unknown $\sin i$ factors (Tabachnik \& Tremaine 2002). 

\subsection{Including the upper limits}\label{sec:avni}

Several methods have been discussed in the literature for finding the distribution most consistent with a set of detections and upper limits (Avni et al.~1980; Feigelson \& Nelson 1985; Schmitt 1985). Such data are known as censored data, and the analysis as survival analysis. Correcting for the upper limits involves counting which of them usefully constrain the distribution at a given point. Avni et al.~(1980) present a method for doing this counting for a one-dimensional distribution. Here, we generalize this approach to the two-dimensional mass-period plane. We follow Avni et al.~(1980) and present a heuristic derivation in this section; a more detailed derivation by a maximum likelihood method is given in the Appendix. The reader may find it useful to compare our discussion with \S III.D of Avni et al.~(1980) which describes the 1-dimensional case.

It is instructive to consider a hypothetical example to develop some intuition. Imagine that after looking at a given star there were three possible outcomes: we either (i) detect a planet, (ii) completely exclude the presence of a planet, or (iii) are not able to say anything (i.e.~cannot exclude or confirm the presence of a planet). First, we observe $N_\star$ stars with the result that $N_{\rm planet}$ planets are found and we are able to rule out planets from all of the remaining stars. The best estimate of the fraction of stars with planets is then $f=N_{\rm planet}/N_\star$. However, what if we are able to rule out planets only from a subset of stars, $N_{\rm no\ planet}<N_\star-N_{\rm planet}$? In this case, the best estimate of the planet fraction is to take the ratio of the number of detections to the number of stars for which a detection was possible, $f=N_{\rm planet}/(N_{\rm no\ planet}+N_{\rm planet})$. We must take $N_{\rm no\ planet}+N_{\rm planet}<N_\star$ as the denominator because for each detected planet, the actual pool of target stars is less than the total pool, because the data for some of the stars are inadequate to detect that planet. As an extreme case, consider looking at 100 stars, and being able to say that one has a planet, one does not, and nothing about the remaining 98 stars ($N_{\rm planet}=1$ and $N_{\rm no\ planet}=1$). The best estimate for the planet fraction is then 50\%. The extra 98 stars for which no useful upper limit could be obtained do not contribute to the estimate. This simple example tells us how to interpret Figure 8. In a region of the mass-period plane in which planets can be ruled out for a fraction $k$ of stars, the best estimate of the number of planets is $\approx 1/k$ times the number of detections.

Our approach is to assign an effective number $N_i$ for each detected planet $i$. If selection effects are unimportant, $N_i=1$, so that the detected planet is a good representation of the number of planets at that mass and period. However, $N_i$ will be greater than 1 if planet $i$ has orbital parameters in part of the mass-period plane where completeness corrections are important. For instance, if the completeness for planets at a given mass and period is 50\%, then $N_i$ for a planet $i$ at that mass and period will be 2. The idea behind this method is that we are sampling the mass and period distribution at the discrete set of points corresponding to the mass and periods of the detected planets. Of course, the underlying distribution is likely to be smooth, and so the quantity $N_i/N_\star$ should be thought of as the probability that a star has a planet with mass and period close to those of planet $i$. The normalization of $N_i$ is such that the total fraction of stars with planets is $\sum_i N_i/N_\star$.

To calculate $N_i$ for a given star with a detected planet, we must count how many of the stars with non-detections could have an undiscovered planet with the same mass and period as planet $i$. We therefore consider each non-detection in turn and ask whether the upper limit $K_{\rm up}$ calculated in \S 2.5 allows a companion with period $P_i$ and $K_i$ to be present. If so, we must increase $N_i$ to allow for this incompleteness. However, the upper limit $K_{\rm up}$ allows a hypothetical planet to be present with {\em any} amplitude or orbital period that satisfies $K<K_{\rm up}$, not just the orbital parameters of planet $i$. Therefore, rather than increasing $N_i$ by 1 for every upper limit that allows additional undiscovered planets at that location, we must increase $N_i$ by the probability that the hypothetical planet would have the same parameters as planet $i$. In effect, we ``share out'' the hypothetical planet amongst all the possible locations in mass period space that are allowed by the upper limit. This leads to the following rule for increasing $N_i$ each iteration:
\begin{equation}\label{eq:iterate}
N_i^{n+1}=1 + \sum_{j,K_{{\rm up}, j}>K_i}\ {N^n_i\over N_\star Z(K_{{\rm up},j})},
\end{equation}
where $n$ labels the iteration and the sum is over all non-detections $j$ which allow an undiscovered planet with the properties of planet $i$ to be present. Here, for each of the non-detections considered in the sum, $K_{{\rm up},j}$ is the upper limit on the velocity amplitude (\S 2.5) at the orbital period of planet $i$. The second term in equation (\ref{eq:iterate}) is the probability that a planet with a velocity amplitude below the upper limit has the period and amplitude corresponding to planet $i$. It is weighted by the normalization factor $Z$ in the denominator which counts all the possibilities consistent with the upper limit: these are either (i) a planet is present with $K<K_{\rm up}$, or (ii) no planet is present. Mathematically, we write this as the probability that the star does not have a planet with an amplitude that violates the upper limit ($K>K_{\rm up}$),
\begin{equation}
Z(K_{\rm up})=1-\sum_{i,\ K_i > K_{\rm up}}{N^n_i\over N_\star},
\end{equation}
where the sum is over all detected planets $i$ whose velocity amplitude $K_i$ exceeds $K_{\rm up}$.

\begin{deluxetable*}{lllllll}
\tablecaption{Cumulative percentage of stars with a planet\tablenotemark{a}\label{tab:cumulative}}
\tablehead{\colhead{} & \colhead{$P<11.5\ {\rm d}$} & \colhead{$100\ {\rm d}$} & \colhead{$1\ {\rm yr}$} & \colhead{$2.8\ {\rm yr}$} & \colhead{$5.2\ {\rm yr}$} & \colhead{$11.2\ {\rm yr}$}\nl
\colhead{} & \colhead{} & \colhead{} & \colhead{} & \colhead{(1022 d)} & \colhead{(1896 d)} & \colhead{(4080 d)}\nl
\colhead{} & \colhead{$a<0.1\ {\rm AU}$\tablenotemark{b}} & \colhead{ 0.42 AU } & \colhead{1 AU} & \colhead{2 AU} & \colhead{3 AU} & \colhead{5 AU}}
\startdata
$M>2\ M_J$ & 0 & 0.43 (0.3) & 1.1 (0.5) & 1.9 (0.6) & 2.6 (0.7) & 4.2 (0.9)\nl
& 0 & 0.42 (0.3) & 1.1 (0.5) & 1.9 (0.6) & 2.5 (0.7) & 4.0 (0.9)\nl
& 0 & 0.45 (0.3) & 1.1 (0.5) & 2.1 (0.7) & 2.8 (0.8) & 3.0 (0.8)\nl
& 0 & 0.42 (0.3) & 1.1 (0.5) & 1.9 (0.7) & 2.5 (0.8) & 2.8 (0.8)\nl
&&&&&&\nl
$1\ M_J$ & 0.43 (0.3) & 0.85 (0.4) & 1.9 (0.6) & 3.9 (0.9) & 4.6 (1.0) & 8.9 (1.4)\nl
& 0.42 (0.3) & 0.85 (0.4) & 1.9 (0.6) & 3.8 (0.9) & 4.4 (1.0) & 8.3 (1.4) \nl
& 0.46 (0.3) & 0.9 (0.4) & 2.1 (0.7) & 4.3 (1.0) & 5.0 (1.0) & 6.3 (1.2) \nl
& 0.42 (0.3) & 0.85 (0.4) & 1.9 (0.7) & 3.8 (1.0) & 4.4 (1.0) & 5.5 (1.2) \nl
\nl
$0.5\ M_J$ & 1.1 (0.5)&  1.9 (0.6) & 3.3 (0.8) & 6.3 (1.2) & 7.5 (1.3) &\nl
& 1.1 (0.5) & 1.9 (0.6) & 3.2 (0.8) & 5.9 (1.2) & 7.0 (1.3) &\nl
& 1.2 (0.5) &  2.1 (0.7) & 3.5 (0.9) & 6.2 (1.2) & 7.2 (1.2) &\nl
& 1.1 (0.5) & 1.9 (0.7) & 3.2 (0.9) & 5.5 (1.2) & 6.4 (1.2) &\nl
\nl
$0.3\ M_J$ & 1.5 (0.6) & 2.4 (0.7) & 3.7 (0.9) & 6.7 (1.2) & 8.5 (1.3) & \nl
& 1.5 (0.6) & 2.3 (0.7) & 3.6 (0.9) & 6.4 (1.2) & 7.6 (1.3) &\nl
& 1.6 (0.6) & 2.6 (0.7) & 4.0 (0.9) & 6.7 (1.2) & 7.7 (1.3) & \nl
& 1.5 (0.6) & 2.3 (0.7) & 3.6 (0.9) & 5.9 (1.2) & 6.8 (1.3) &\nl
\nl
$0.1\ M_J$ & 2.0 (0.7) & 3.9 (0.9) &   & & & \nl
& 1.9 (0.7) & 3.4 (0.9) & & & & \nl
& 2.2 (0.7) & 4.3 (1.0) & & & & \nl
& 1.9 (0.7) & 3.4 (1.0) & & & & 
\enddata
\tablenotetext{a}{The percentage of F,G, and K stars with a planet at a shorter orbital period than the one given, and with $M_P\sin i$ greater than the given mass, but less than $15\ M_J$. In each case, we give four values: (1) the percentage derived by including both announced and unannounced detections, (2) the same as (1) but without completeness corrections, (3) the percentage derived from including the announced planets as detections but treating the unannounced detections as upper limits, (4) same as (3) but without completeness corrections. The total number of stars is 475. The Poisson error is given in parentheses after each entry. See \S 2.1 for a description of the stellar sample selection and distribution of spectral types.}
\tablenotetext{b}{The corresponding semimajor axis for a solar mass star.}
\end{deluxetable*}

For example, suppose we wish to calculate $N_i$ for planet $j$ with $K=20\ {\rm m\ s^{-1}}$. We start with our initial guess of $N^0_i=1$ for all $i$. We then turn to planet $j$, and, for that planet, consider each star with a non-detection in turn. Let us say that the first such star has high jitter or a small number of data points, so that the upper limit on the velocity amplitude of a companion is  $30\ {\rm m\ s^{-1}}$. This means that a companion with an amplitude of $20\ {\rm m\ s^{-1}}$, the same as planet $j$, cannot be ruled out, and so we must add a contribution to $N_j$. To do this, we first use equation (10) to calculate $Z(30\ {\rm m\ s^{-1}})$, where the sum is over all detected planets with $K>30\ {\rm m\ s^{-1}}$. If there are 30 such planets, and 500 total stars, then using the current values $N^0_i=1$ (this is the first iteration), we find $Z(30\ {\rm m\ s^{-1}})=1-30/500=0.94$. This is the probability (given our current values for $N_i$) that a star does not have a planet with $K>30\ {\rm m\ s^{-1}}$. We use this value in the first term of the sum in equation (9), which means that we add an amount $(1/500)(1/0.94)=2.1\times 10^{-3}$ to $N^0_j$. Because we know that there is no planet with $K>30\ {\rm m\ s^{-1}}$ around this star, the probability that there is a planet with the properties of planet $j$ is larger than $1/500$. We now continue with the sum. Let us say that the second star with a non-detection is well-observed and those observations rule out a planet with the properties of planet $j$. This star then contributes nothing to the sum in equation (9).  We continue for all stars with non-detections to complete the sum in equation (9) and arrive at the new value $N^1_j$. We then repeat this calculation to obtain $N^1_i$ for all detected planets $i$. This entire procedure is iterated until all values of $N_i$ have converged.

An important question is how much our results depend on the candidate detections, since these detections await further observations before they can be confirmed as being due to an orbiting planet. Long period signals need further observations to cover at least one orbit; low amplitude signals are potentially due to other factors such as stellar jitter. Therefore it is likely that some of these candidate periodicities are not due to a planet and should not be included. Even if they are due to a planet, as more data are collected, the best-fit orbital parameters of these candidates may change. To answer this question, we have calculated the values of $N_i$ both with and without the candidate detections, by either including the candidate detections or by treating them as non-detections, with the upper limit $K_{\rm up}$ given by the detected velocity amplitude. We find that our conclusions regarding the mass-orbital period distribution are similar in each case (see, for example Table \ref{tab:cumulative} discussed below).

\subsection{A power law fit using maximum likelihood}\label{sec:power}

As an alternative to the non-parametric description of the period and mass distribution that we described in the last section, we also fit the distribution with the simplest parametric model, a power law in mass and period. One way to do this would be to take the corrected data given by the $N_i$ values we have described, and fit a power law to the histogram or the cumulative distribution. Instead, we have used a maximum likelihood technique to fit a power law in mass and orbital period simultaneously to the original data (see also Tabachnik \& Tremaine 2002). In the Appendix we start with the same likelihood that is used to derive the non-parametric technique described in \S 3.1, but instead consider a parametric model in which the probablility of a star having a planet at mass $M$ and period $P$ is $dN=C M^\alpha P^\beta\ d\ln M\ d\ln P$, where $C$ is a normalization constant. The resulting expression for the likelihood is given in equation (\ref{eq:like3}), and we evaluate this numerically and maximize it with respect to the two parameters $\alpha$ and $\beta$. 

\subsection{Results for F, G, K dwarfs}

We first present results for the 475 stars with $M_\star>0.5\ M_\odot$ which have F, G, and K spectral types (see \S 2.1 for a discussion of the sample selection and the range of spectral types). We do not attempt a detailed study of the stellar mass dependence of planet occurrence rate or orbital parameters in this paper. Fischer \& Valenti (2005) show that the apparent rate of occurrence of planets increases by a factor of 2 for stellar masses between $0.75$ and $1.5 M_\sun$. Investigating the stellar mass dependence of planet properties requires untangling it from the effect of stellar metallicity, beyond the scope of this paper (e.g., see the recent discussion in Johnson et al.~2007). However, several recent papers have pointed out that the planet occurrence rate around M dwarfs appears to be several times lower than around F, G and K stars (Butler et al.~2004b, 2006b; Endl et al.~2006; Johnson et al.~2007). To avoid biasing our results for the occurrence rate of planets, and to investigate the occurrence rate around M dwarfs further, we treat stars with masses $<0.5\ M_\sun$ separately in \S \ref{sec:Mdwarfs}. In addition, three of the 48 stars with announced planets were added to the survey to confirm detections by other groups. To ensure a fair sample, we remove these stars from our analysis. 

Figure \ref{fig:avnim} shows the results of the calculation described in \S \ref{sec:avni}. We plot the mass and period of announced planets (black circles) and candidate detections (green circles), as well as the upper limit contours (blue curves) as in Figure \ref{fig:keckm}. The value of $N_i$ for each detection is indicated by the area of the circle. The smallest circles, well above the detection threshold, correspond to $N_i=1$. At lower masses, the values of $N_i$ are roughly consistent with the simple argument given in \S 3.1, that if the detection lies at a mass and period excluded by a fraction $k$ of the upper limits, then roughly $N_i\approx 1/k$. For $K\approx 10\ {\rm m\ s^{-1}}$, the completeness corrections are roughly a factor of two, and are close to unity for $K\gtrsim 20\ {\rm m\ s^{-1}}$. 

\begin{figure}
\epsscale{1.0}\plotone{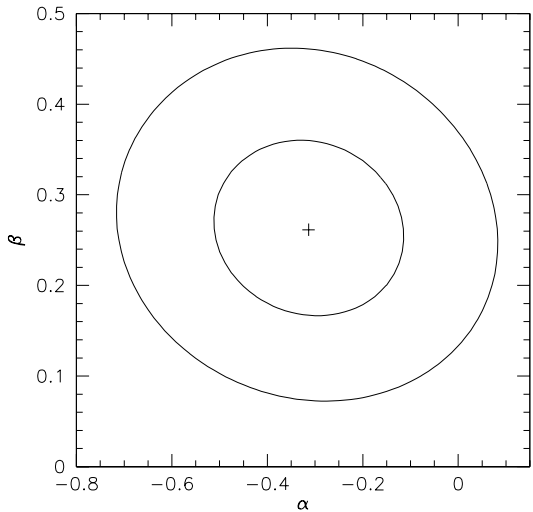}
\caption{Contours of constant likelihood for a power law fit $dN\propto M^\alpha P^\beta\ d\ln M\ d\ln P$. The contours correspond to the 68\% and 95\% confidence intervals ($\Delta L/2=1$ and $4$). The best fitting values are $\alpha=-0.31$ and $\beta=0.26$. We include announced planets and candidates in the period range $2$-$2000$ days and mass range $0.3$--$10\ M_J$ in the fit.
\label{fig:avni4cont}}
\end{figure}

Figure \ref{fig:avni4cont} shows the result of our power-law fit (\S \ref{sec:power}). We fit the data in the mass range $0.3$--$10\ M_J$ and $2$--$2000$ days, which includes 32 announced planets, and 4 candidate detections. The constraints on $\alpha$ and $\beta$ are shown in Figure \ref{fig:avni4cont}. The best fit values which maximize the likelihood are $\alpha=-0.31\pm 0.2$ and $\beta=0.26\pm 0.1$. The error in $\alpha$ is larger than the error in $\beta$, presumably because the dynamic range in orbital periods is greater than in the planet masses. The best fitting power law gives a fraction of stars with a planet of $10.5$\% in this mass and period range, which compares with the value of 8.5\% derived from our completeness corrections (see Table \ref{tab:extrapol}). The values of $\alpha$ and $\beta$ are slightly correlated, in the same sense as Tabachnik \& Tremaine (2002) observed, but to a smaller degree.

\subsubsection{Fraction of stars with a planet}

Summing $N_i$ in a particular region of the mass-orbital period plane and dividing by $N_\star$ gives the fraction of stars with planets in that region. The percentage of stars with a planet more massive than a given mass and closer to the star than a given orbital period is listed in Table \ref{tab:cumulative}. In parentheses we give the Poisson error based on the number of detections. For each table entry, we give four values. The first two are the percentages derived by including both announced planets and unannounced candidates, with and without completeness corrections included. The third and fourth values are the percentages derived by including the announced planets only (in which case the velocity amplitude of each candidate is treated as an upper limit on velocity amplitude rather than a detection), with and without completeness corrections included. The two values are similar over most of Table \ref{tab:cumulative}. The largest differences of $\approx 30$\% are at long periods $>2000$ days, where there are very few announced planets, and many candidate periodicities.

\begin{figure}
\epsscale{1.0}\plotone{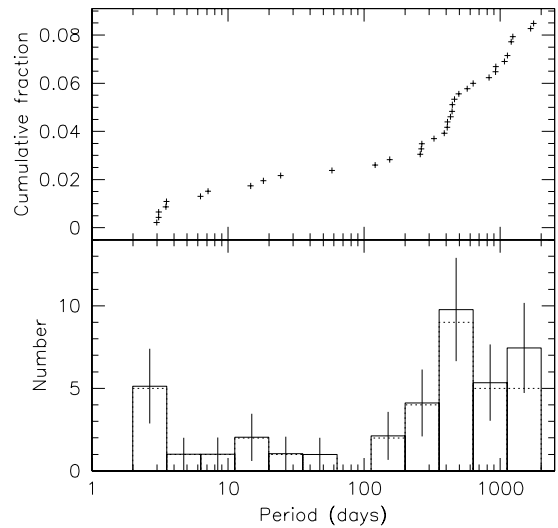}
\caption{Distribution of orbital periods for planets with periods$<2000$ days and mass $M_p\sin i>0.3 M_J$. In the lower panel, the dotted histogram shows the number of detections in each bin, including announced and candidate detections; the solid histogram shows this number corrected for completeness. Error bars indicate $\sqrt{N}$ for each bin. The upper panel shows the cumulative percentage of stars with a planet with orbital period smaller than a given value.
\label{fig:avnihistp4}}
\end{figure}

\subsubsection{Distribution of orbital periods}

Figure \ref{fig:avnihistp4} shows the orbital period distribution for planets with $M_P\sin i>0.3\ M_J$ for orbital periods up to $2000$ days, beyond which the detectability declines as the orbital period approaches the duration of the survey. In the lower panel, the dotted histogram is the distribution of detections, including announced and candidate detections; the solid histogram is the distribution of detections after correcting for completeness, i.e.~summing $N_i$ in each bin. For each bin, we indicate the expected Poisson $\sqrt{N}$ errors based on the number of detections but rescaled by the ratio of $\sum N_i$ in that bin to the number of detections in that bin. The upper panel shows the cumulative histogram, showing the fraction of stars with a planet within a given orbital period. For clarity, we do not show the Poisson errors on the cumulative histogram, but they can be calculated based on the total number of stars. For example, approximately 2.4\% of stars have a planet more massive than $0.3\ M_J$ with an orbital period of $<100$ days. This represents  $11.3\pm 3.4$ stars out of the total of 472, or a fraction $2.4\pm 0.7$\%.

\begin{figure}
\epsscale{1.0}\plotone{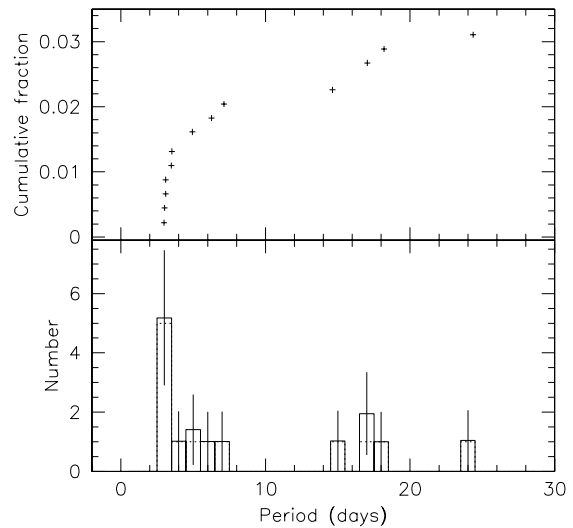}
\caption{Same as Figure \ref{fig:avnihistp4}, but now for short orbital periods $P<30\ {\rm days}$, and for masses $M_P\sin i>0.1\ M_J$. All detections correspond to announced planets in this period and mass range.
\label{fig:avnihistp3}}
\end{figure}

\begin{deluxetable}{lllll}
\tablewidth{0pt}
\tablecaption{Extrapolated occurrence rates of long period orbits\tablenotemark{a}\label{tab:extrapol}}
\tablehead{ \colhead{Model} & \colhead{$P<5.2\ {\rm yr}$} & \colhead{$11\ {\rm yr}$} & \colhead{$32\ {\rm yr}$} & \colhead{$89\ {\rm yr}$}\nl
\colhead{} &  \colhead{$a<3\ {\rm AU}$} & \colhead{$5\ {\rm AU}$} & \colhead{$10\ {\rm AU}$} & \colhead{$20\ {\rm AU}$}
}
\startdata
flat & 8.5 (1.3) & 11 (1.7) & 14 (2.1) & 17 (2.6)\nl
$dN/d\log_{10}P=6.5$\% &&&& \nl
power law & 8.5 (1.3) & 11 (1.7) & 14 (2.3) & 19 (3.0)\nl
$d\ln N/d\ln P=0.26$ &&&&\nl
\enddata
\tablenotetext{a}{The cumulative percentage of stars with a planet $N(<P)$, based on either a flat extrapolation or power law extrapolation beyond $P=2000$ days, for planet masses $0.3<M_P\sin i<15\ M_J$. The semimajor axes are for a solar mass star. The extrapolated Poisson error is indicated in parentheses.}
\end{deluxetable}

At the shortest periods, the period distribution shows the well known pile up of planets at orbital periods close to 3 days. This is illustrated in more detail in Figure \ref{fig:avnihistp3} which shows the period distribution for those planets with $P<30$ days, and masses $M_P\sin i>0.1\ M_J$ (the increased detectability of close in planets means that we can go to lower masses than in Fig.~\ref{fig:avnihistp4}). All of these planets are announced; there are no candidate planets in this range of mass and period. Butler et al.~(1996) mention that there are  no significant selection effects that would lead to this pile up: we can see that very clearly in Figure \ref{fig:avnim}, in which the upper limit curves continue smoothly to periods as short as 1 day with no change in detectability. The statistical significance of the pile up in our data depends on the model and range of orbital periods against which it is compared. A Kolmogorov-Smirnov (KS) test gives a 0.4\% probability that the observed distribution is drawn from a uniform distribution in $\log P$ in the decade $1$ to $10$ days. If we extend the uniform distribution out to 100 days, the KS probability is 20\%.

As we described earlier, a power law fit to the period distribution gives $dN/d\ln P\propto P^{0.26}$, which rises to longer periods. However, an alternative description of the distribution is a rapid increase in the planet fraction at orbital periods of $\approx 300$ days. Figure \ref{fig:avnihistp4} shows that there is a change of slope in the cumulative distribution which suggests that the planet fraction increases beyond orbital periods of $\approx 300$ days. The change in slope does not depend on whether the candidate planets are included. If we assume that the orbital period distribution above and below $300$ days is flat, we find that the fraction of stars with a planet per decade is $dN/d\log_{10}P=1.3\pm 0.4$\% at short periods, and $dN/d\log_{10}P=6.5\pm 1.4$\% at long periods (the latter becomes $dN/d\log_{10}P=5.1\pm 1.2$\% if only announced planets are included). Therefore, the incidence of planets increases by a factor of $5$ for periods longwards of $\approx 300$ days. The low planet fraction at intermediate orbital periods has been noted previously. Jones et al.~(2003) and Udry et al.~(2003) both pointed out that there is a deficit of gas giants at intermediate periods, $P\approx 10$--$100$ days. 

We have focused on the period distribution for $P<2000$ days, but Figure \ref{fig:avnim} shows that there are many candidate planets with orbital periods $P>2000$ days. In our analysis, we have assumed that the minimum mass and orbital period of detected companions are well-determined. However, for orbital periods longer that the time-baseline of the data, this is not the case: there exists a family of best-fitting solutions with a range of allowed orbital periods, masses, and eccentricities (see e.g.~Ford 2005; Wright et al.~2007). To constrain the distribution at long orbital periods requires taking into account the distributions of orbital parameters allowed by the data for each star. For now, we extrapolate the period distribution determined for $P<2000$ days to predict the occurrence rate of long period orbits assuming that either the flat distribution or the power law $\propto P^\beta$ holds for longer orbital periods. The results are given in Table \ref{tab:extrapol}. For example, if the distribution is flat in $\log P$ beyond $2000$ days, we expect that 17\% of solar type stars harbor a gas giant (Saturn mass and up) within 20 AU. In the power law case, the fractions are larger, but not substantially larger because of the small value of $\beta=0.26$. These extrapolations are consistent with the number of candidates we find at long orbital periods. If we sum the confirmed planets and candidates at long periods, taking the completeness corrections into account, and assuming that the fitted orbital periods are the correct ones, we find that 18\% of stars have a planet or candidate within $10\ {\rm AU}$. This is the same fraction as our extrapolations suggest for orbital separations less than $20\ {\rm AU}$. However, the uncertainties in orbital parameters need to be taken into account before we can use the long period candidates to learn about the period distribution at long orbital periods.

\begin{figure}
\epsscale{1.0}\plotone{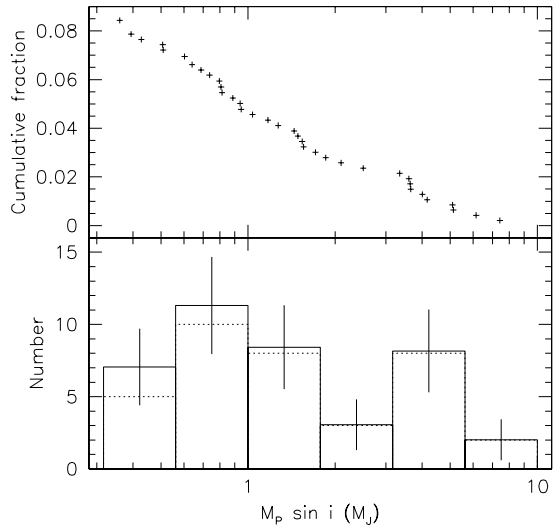}
\caption{The distribution of $M_p\sin i$ for planets with $P<2000$ days and $M_P\sin i>0.3\ M_J$. In the lower panel, the dotted histogram shows the number of detections in each bin; the solid histogram shows the number corrected for completeness. The cumulative distribution is shown in the upper panel.\label{fig:avnihistm3}}
\end{figure}

\subsubsection{The mass function of planets}

The $M_P\sin i$ distribution of planets with $M_P\sin i>0.3\ M_J$ and $P<2000$ days is shown in Figure \ref{fig:avnihistm3}. As we have discussed, a power law fit to the mass function gives $dN/d\ln M\propto M^{\alpha}$, with $\alpha=-0.31\pm 0.2$, so that the distribution in $\ln M$ is approximately flat, but slowly rising to lower masses. Our value for $\alpha$ agrees with the $dN/dM\propto M^{-1.1}$ found by Butler et al. (2006a) by fitting the mass function of planets detected in the combined Keck, Lick, and AAT surveys (see also Jorissen, Mayor, \& Udry 2001). The cumulative distribution in the upper panel of Figure \ref{fig:avnihistm3} (which shows the fraction of stars with a planet more massive than a given mass) shows a corresponding close to linear increase to lower masses. The absence of a turnover in the cumulative distribution at low masses shows that our results are consistent with the mass function remaining approximately flat in $\ln M$ to the lowest masses with good detectability.

\begin{figure}
\epsscale{1}\plotone{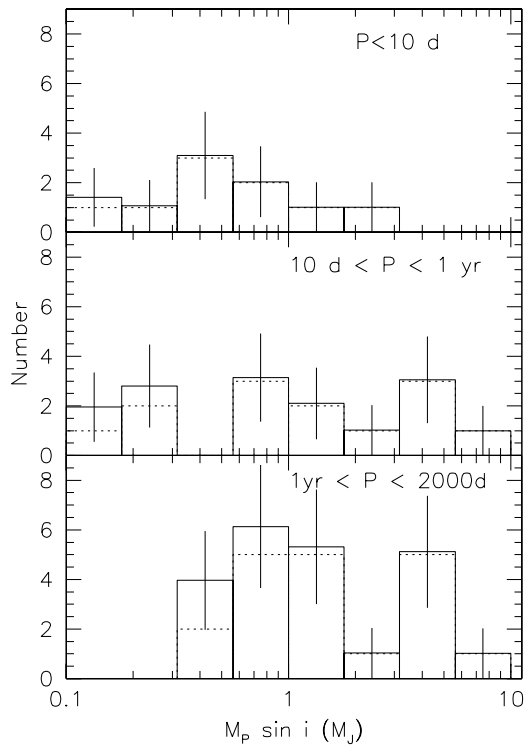}
\caption{Histogram of $M_p\sin i$ for planets in different period ranges. The dotted histogram shows the number of detections in each bin; the solid histogram shows the number corrected for completeness. \label{fig:avnihistm2}}
\end{figure}

\begin{figure}
\epsscale{1}\plotone{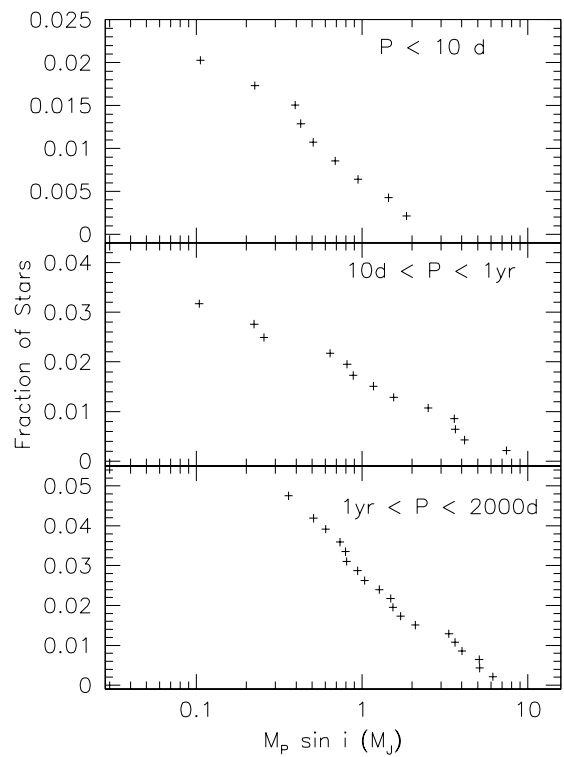}
\caption{Cumulative distribution of $M_p\sin i$ for planets in different period ranges. \label{fig:avnihistm}}
\end{figure}

An important question is whether the mass function is dependent on orbital period. In Figures \ref{fig:avnihistm2} and \ref{fig:avnihistm}, we show the mass distributions in three different period ranges: periods less than $10$ days, between $10$ days and 1 year, and greater than 1 year. In the context of theoretical models for planet formation and migration, these ranges correspond to planets that have undergone different amounts of migration (e.g.~Ida \& Lin 2004). For orbital periods less than a year, we give the distribution down to $0.1\ M_J$, but for longer orbital periods where detectability is not as good, we restrict the mass range to $>0.3 M_J$. We detect no close in planets with $M>2 M_J$. This lack of close, massive planets has been noted before, and is significant: Figure \ref{fig:avnim} shows that the survey is complete in that region of the mass-period plane. At the longest orbital periods, there are almost as many detections in the mass range $0.3<M_P\sin i<1$ (half a decade) as there are in the range $1<M_P\sin i<10$ (a full decade), suggesting a steeper fall off with mass than the overall mass distribution. However, this depends on future confirmation of the candidates with additional observations, and depends on the completeness corrections, which are significant in the lowest mass bin.

Ida \& Lin (2004) predict a mass-orbital period ``desert'' at low masses and intermediate orbital periods. Our data show no evidence for a drop in the planet occurrence rate at low masses, as can be seen in the central panel of Figure \ref{fig:avnihistm2}, although as Figure \ref{fig:avnim} shows, we are not able to address the distribution for masses $\lesssim 0.1$--$0.2\ M_J$ in this period range because of selection effects. In their study of the mass-period distribution, Udry et al.~(2003) found evidence for a deficiency of planets with $M_P\sin i<0.75\ M_J$ at orbital periods $\gtrsim 100$ days. They simulated the detectability of planets in that region, and concluded that detections would have been made if the mass distribution of planets at $P>100$ days was similar to that of the hot jupiters. Interestingly, we find Saturn mass candidates at periods longer than 1000 days, but not in the range $100$--1000 days. However, the small number of detections in that region of the mass-period plane and the fact that the completeness corrections are significant there mean that we cannot come to any definite conclusions.

Fitting a uniform distribution in $\log M$ to the distributions in Figure \ref{fig:avnihistm2}, we find a fraction per decade of $dN/d\log_{10}M=1.8\pm 0.6$\% for  $P<10\ {\rm d}$, $1.9\pm 0.5$\% for $10\ {\rm d}<P<1\ {\rm yr}$, and $3.9\pm 0.9$\% for $1\ {\rm yr}<P<2000\ {\rm days}$. It is interesting to extrapolate this distribution to lower masses. For example, at short periods $P<10$ days, we expect 3.1\% of stars to have a planet more massive than 10$M_\Earth$, and $4.7$\% to have an Earth mass planet or larger. For periods less than 1 year, extrapolation gives 7.4\% of stars with $M_P>10\ M_\Earth$, and 11\% with $M_P>1\ M_\Earth$.

\subsubsection{Comparison with previous work}

Our results for the fraction of stars with planets and the mass and orbital period distributions are generally consistent with previous determinations. Marcy et al.~(2005a) estimate that 12\% of stars have a gas giant within 20 AU, based on a flat extrapolation of the orbital period distribution of the detected planets in the combined Lick, Keck, and Anglo-Australian surveys. If we take announced planets only without completeness corrections, we find that the fraction of stars with a planet per decade is $4.8\pm 1.0$\%, which extrapolates to $13\pm 2.5$\% within 20 AU in good agreement with Marcy et al.~(2005a).

Tabachnik \& Tremaine (2002) fit the mass and period distributions with a double power law, $dN\propto M^\alpha P^\beta d\ln M d\ln P$, accounting for selection effects by estimating the detection thresholds for each of the Doppler surveys. They obtained $\alpha=-0.11\pm 0.1$, which agrees with our value $\alpha=-0.31\pm 0.2$, and $\beta=0.27\pm 0.06$, which agrees with our $\beta=0.26\pm 0.1$. Our error bars are larger than those of Tabachnik \& Tremaine (2002) because in their work they included several surveys, and so have a larger total number of stars in their sample (their results for the Keck survey alone have similar error bars to our results). Extrapolating, Tabachnik \& Tremaine (2002) found that 4\% of solar type stars have planets with $1 M_J<M<10 M_J$ and $2\ {\rm d}<P<10\ {\rm yr}$. Our results give 6--9\% depending on how the candidate detections are included (Table \ref{tab:cumulative}).

Lineweaver \& Grether (2003) also fit a double power law in mass and period. Their technique was to define an area in which they estimate all planets have been detected around stars currently being surveyed, and extrapolate to longer periods and lower masses. For the mass function, they found $\alpha=-0.8\pm 0.3$, i.e.~a significant rise in the mass function at low masses. We do not observe such a rise in Figure~\ref{fig:avnihistm3}. For the period distribution, they found $\beta=0.7\pm 0.3$ which again indicates a rise in the planet occurrence rate at long periods. They estimated that 9\% of stars have a planet with $M>0.3 M_J$ and $P<13\ {\rm yrs}$. Our extrapolation suggests a fraction $11$\% in this mass and period range (see Table \ref{tab:extrapol}). 

Naef et al.~(2005) present an analysis of data for 330 stars with 18 detected planets from the ELODIE Planet Search program. They give the fraction of stars with planets more massive than $0.5 M_J$ within three different orbital periods: $0.7\pm 0.5$\% for $P<5\ {\rm d}$, $4.0\pm 1.1$\% for $P<1500\ {\rm d}$, and $7.3\pm 1.5$\% for $P<3900\ {\rm d}$. For the same mass and period ranges, we find $0.65\pm 0.4$\%, $6.9\pm 1.2$\% and $12\pm 1.6$\% (this last number is $8.6\pm 1.3$\% if only announced planets are included).

\subsection{Results for M dwarfs}
\label{sec:Mdwarfs}

\begin{figure}
\epsscale{1.0}\plotone{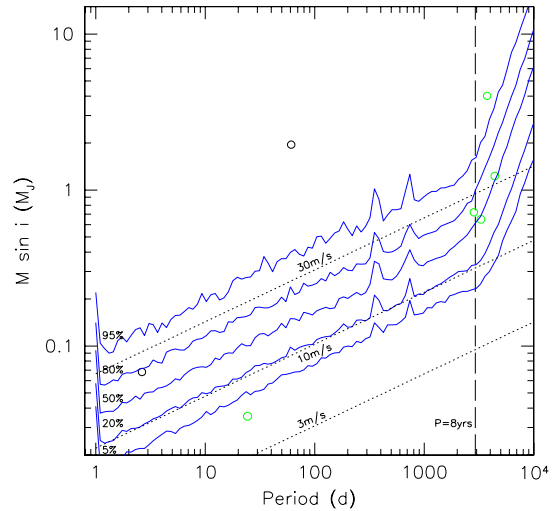}
\caption{Same as Figure \ref{fig:avnim}, but showing only the detections for the 110 stars with $M_\star<0.5 M_\odot$. The blue curves show the upper limits corresponding to this mass range. The dotted lines show velocity amplitudes of $3,10$ and $30\ {\rm m\ s^{-1}}$ for a $0.3\,M_\odot$ star.\label{fig:avnimM}}
\end{figure}

We now turn to the M dwarfs in the sample. Figure \ref{fig:avnimM} shows the results of the calculation described in \S \ref{sec:avni} applied to the 110 stars with $M<0.5\ M_\sun$. Having detected 46 planets from 475 F, G, and K stars, we would expect 11 planets in this sample of M dwarfs if the planet frequency and detectability were the same. Instead, there are only 2 announced planets: GJ 876, which is in fact a triple system but our search detects only the most massive planet at $P=60$ days, and GJ 436, a $0.07\ M_J$ planet in an orbit of less than 3 days.

Several recent papers have addressed the apparent paucity of gas giant planets around M dwarfs. In their paper announcing the discovery of the Neptune-mass planet orbiting GJ 436, Butler et al.~(2004b) estimated that the planet fraction for M dwarfs is $\approx 0.5$\% for masses $\gtrsim 1\ M_J$ and periods $<3$ years, roughly an order of magnitude lower than around F and G main sequence stars. Butler et al.~(2006b) announced the detection of a $0.8\ M_J$ planet in an orbit with $P=1890$ days around GJ 849. Including both GJ 876 and GJ 849, they estimate a planet fraction $2/114=1.8\pm 1.2$\% for planet masses $\gtrsim 0.4\ M_J$ and periods $a<2.5$ AU. In the same range of $a$ but with a larger mass limit $M_P\sin i>0.8 M_J$, Johnson et al.~(2007)  find that this fraction is $1.8\pm 1.0$\% for stars in the mass range $0.1$-$0.7 M_\odot$. Endl et al. (2006) give a limit on the fraction of M dwarfs with planets of $<1.27$\% at the $1\sigma$ confidence level. This result is less constraining, since they estimate that their survey of 90 M dwarfs is 95\% complete for $M_P\sin i>3.5\ M_J$ and $a<0.7$ AU, and Table \ref{tab:cumulative} shows that we find a planet fraction of $\approx 1$\% for planets in this mass and period range around FGK stars. These observational constraints agree with predictions of core accretion models for planet formation, which find that Jupiter mass planets should be rare around M dwarfs, with the mass function of planets shifted towards lower masses (Laughlin et al.~2004; Ida \& Lin 2005; Kennedy \& Kenyon 2008) (although Kornet \& Wolf 2006 predict a greater incidence of gas giants at lower stellar masses if the protoplanetary disk parameters do not change with stellar mass).

\begin{figure}
\epsscale{1.0}\plotone{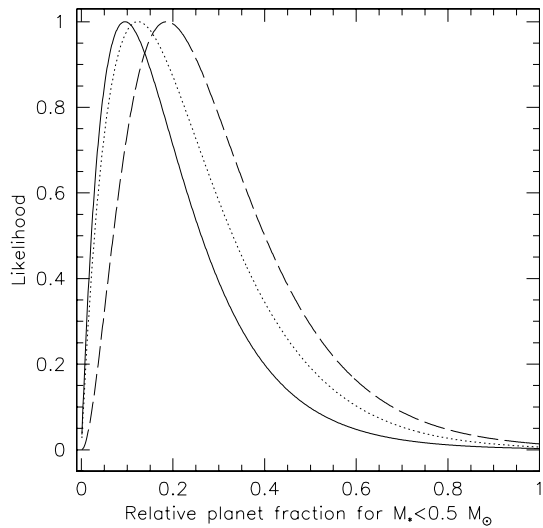}
\caption{Relative fraction of stars with a planet for stars with $M_\star<0.5\ M_\odot$ compared with those with $M_\star>0.5\ M_\odot$. For each set of stars, we assume the mass-period distribution is $dN=C M^\alpha P^\beta d\ln Md\ln P$ with $\alpha=-0.31$ and $\beta=0.26$ (the best fit values for the whole sample, see Fig.~\ref{fig:avni4cont}) and plot the likelihood of the ratio of the normalization factors $r$. The dotted curve shows the expected distribution if selection effects are not important, i.e.~ based on Poisson counting statistics only. The dashed curve shows the result if we include the recently announced companion to GJ~849.\label{fig:avni5}}
\end{figure}

With so few detections in our sample we obviously cannot say anything about the mass-period distribution. However, we can address the possible role of selection effects and constrain the relative fractions of stars with planets around $M$ dwarfs compared to FGK stars. We again take a maximum likelihood approach. We assume that the mass-period distribution for M dwarfs is the same power law distribution that we fit for the larger sample of FGK stars, $dN=C\ M^\alpha P^\beta\ d\ln M\ d\ln P$, with $\alpha=-0.31$ and $\beta=0.26$, but with a different normalization constant $C$. (The mass-period distribution is likely different for M dwarfs than FGK stars, but this is the simplest assumption given the available data). We then calculate the likelihood for the ratio of normalization constants $r=C(M_\star<0.5\ M_\sun)/C(M_\star>0.5\ M_\sun)$. This is shown in Figure \ref{fig:avni5}. 
We use the same mass range $0.3$--$10\ M_J$, and period range $2$--$2000$ days as in \S 3.1. The best fitting value is $r=0.095$, indicating that M dwarfs are 10 times less likely to harbor a gas giant within 2000 days. The 95.4\% ($2\sigma$) upper limit on the relative planet fraction is 0.51. Using the normalization from the power law model (which for the best-fitting model has 10.5\% of stars with planets for FGK stars), we find the best-fitting M dwarf planet fraction to be $1.0$\%, with a $2\sigma$ upper limit of $5.4$\%. 

The shape of the curve in Figure \ref{fig:avni5} is straightforward to understand. In the period range $P<2000$ days and mass range $0.3$--$10\ M_J$, there are 35 detections out of 475 FGK stars, a fraction of 7.4\%. If the planet fraction around M dwarfs is $r$ times the planet fraction around FGK dwarfs, we therefore expect to find $8.1r$ detections amongst the 110 M dwarfs. The probability of detecting 1 planet is then $(8.1r)\exp(-8.1r)$ from the Poisson distribution. Using Bayes' theorem (e.g.~Sivia 1996), we can view this expression as the probability distribution function for $r$ given the measured number of detections. We plot this as the dotted curve in Figure \ref{fig:avni5}. The fact that the dotted curve lies to the right of the maximum likelihood result indicates that the selection effects favor detection of gas giants around M dwarfs by about 25\% (if we increase the expected number by 25\%, from $8.1r$ to $10r$, the solid and dotted curves lie almost on top of each other). Although the Doppler errors and upper limits on velocity amplitude are generally greater for the M dwarfs, the lower stellar mass means that a gas giant planet intrinsically gives a larger velocity signal. The net effect is a slightly larger detectability of gas giants for M dwarfs.

This result shows that the deficit of gas giants around M dwarfs is statistically significant in our sample, and is not due to selection effects against finding companions to M dwarfs. However, the best-fitting value of the ratio $r$ is subject to small number statistics (Fig.~\ref{fig:avni5}). An illustration of this is the recently announced companion to GJ 849 (Butler et al.~2006b), which is one of the long period candidates shown in Figure \ref{fig:avnimM}. The data we analyse here do not include recent observations of this star taken in 2005 and 2006 that show the closure of the orbit. As a result, our search algorithm finds an orbital period of 4400 days, more than twice as long as the true orbital period of $1890$ days. Therefore GJ~849 is actually just inside the region considered above ($P<2000$ days), suggesting that the best current estimate for the M dwarf planet fraction is $\approx 2$\% (Butler et al.~2006b) within 2000 days. The dashed curve in Figure \ref{fig:avni5} shows the result if we include the companion to GJ~849 in our calculation (by correcting the orbital period to the announced value by hand). Again, our result is well approximated by a Poisson distribution (but this time with two detections) if the expected number is increased by 25\%. The best fit relative fraction is $r=0.19$, which corresponds to $2.0$\% of M dwarfs with gas giants within 2000 days. The $2\sigma$ confidence limit on $r$ is $r<0.65$.

\section{Summary and Conclusions}

We have carried out a systematic search for planets using precise radial velocity measurements of 585 stars from the Keck Planet Search. The number, duration, and frequency of observations, and typical Doppler measurement errors are summarized in Figures~\ref{fig:obs1} and \ref{fig:obs2}. This analysis provides a snapshot of the Keck Planet Search at the time of the HIRES spectrometer upgrade in 2004 August.

We systematically searched for planets by calculating the false alarm probability associated with Keplerian orbit fits to the data for each star (C04; Marcy et al.~2005b; O'Toole et al.~2007). This method allows the detection threshold for each star to be understood in terms of the number and duration of the observations, and the underlying ``noise'' from measurement errors, intrinsic stellar jitter, or additional low mass planets. The results are summarized in Figure~\ref{fig:keck2}. We find that all planets with orbital periods $P<2000$ days, velocity amplitudes $K>20\ {\rm m\ s^{-1}}$, and eccentricities $e\lesssim 0.6$ have been announced. For stars without a detection, upper limits (Fig.~\ref{fig:up}) are typically $10\ {\rm m\ s^{-1}}$ for orbital periods less than the duration of the observations, and increase $\propto P^2$ for longer periods (see C04 for a discussion of the period dependence of the detection threshold). The upper limits constrain the presence of additional planets, and allow us to study the mass and orbital period distribution. In section 3, we described a method to calculate the completeness corrections to the mass-orbital period distribution at low masses and long orbital periods. Our method is a generalization of the iterative method of Avni et al.~(1980) to two dimensions. In the Appendix, we show that our approach corresponds to a maximum likelihood method with simple approximations for the likelihood functions of detections and non-detections.

The resulting completeness corrections for the 475 F, G and K stars in the sample are summarized in Figure \ref{fig:avnim}, and Table~\ref{tab:cumulative} gives the fraction of stars with a planet as a function of minimum mass and orbital period (see \S 2.1 for details of the stellar sample including the distribution of spectral types). For masses $>0.3\ M_J$, the detectability is good for periods as large as 2000 days. A power law fit to the data in this range gives a mass-period distribution $dN=C\ M^\alpha P^\beta d\ln Md\ln P$ with $\alpha=-0.31\pm 0.2$ and $\beta=0.26\pm 0.1$. The normalization constant $C$ is such that the fraction of FGK stars with a planet in the mass range $0.3$--$10\ M_J$ and period range $2$--$2000$ days is 10.5\%. In units corresponding to measuring planet masses in Jupiter masses and orbital periods in days, the value of the normalization is $C=1.04\times 10^{-3}$.

Table \ref{tab:extrapol} shows the expected planet fractions obtained by extrapolating our results out to long periods. We estimate that $17$--$19$\% of stars have a gas giant planet within 20 AU. Extrapolating to low masses gives 11\% of stars with an Earth mass planet or larger within 1 AU.  This extrapolation is uncertain, since it takes the distribution derived for gas giant planets into the mass range of rocky planets, for which the formation and migration history is presumably quite different, e.g. Ida \& Lin (2004a). A similar uncertainty applies for our extrapolation to long orbital periods also, since for example the relevant timescales for planet formation grow longer at larger orbital radii, although outwards migration can populate long period orbits (Veras \& Armitage 2004; Martin et al.~2007).

We find several interesting features in the mass-period distribution. Massive planets ($\gtrsim 2\  M_J$) are rare at short orbital periods, as has been noted previously. There is no significant evidence for a lower cutoff in the mass function at intermediate orbital periods, down to planet masses of $0.1$--$0.2\ M_J$. Therefore we do not see any evidence yet for the planet desert proposed by Ida \& Lin (2004a). For orbital periods longer than a year, there are almost as many detections in the mass range $0.3<M_P\sin i<1$ (half a decade) as there are in the range $1<M_P\sin i<10$ (a full decade), suggesting a steeper fall off with mass than the overall mass distribution at long periods. However, this result depends on several candidate detections with $\lesssim M_J$ in this period range that await confirmation. A dependence of the mass function on orbital period might indicate differences in migration mechanisms for different planet masses (Armitage 2007). The orbital period distribution shows an increase in the occurrence rate of gas giants of a factor of 5 beyond $P\approx 300$ days. Theoretical models of planet formation generally predict a smooth increase in the incidence of gas giants at longer orbital periods due to the increasing rate of migration as a planet moves inwards through the protoplanetary disk (e.g.~Figure 1 of Ida \& Lin 2004b; Figure 5 of Armitage 2007). The sharp increase in the period distribution at $P\approx 300$ days shown in Figure 11 may reflect some particular radial structure in the protoplanetary disk. Ida \& Lin (2008b) propose an explanation for this upturn based on a surface density enhancement at the ice line due to a local pressure maximum in the disk. Detailed comparisons between these various models and our results would constrain parameters such as the ratio of migration to disk depletion timescales (e.g.~Ida \& Lin 2004; Armitage 2007).

Because of the small number of detections in the sample of 110 M dwarfs, we are not able to constrain the mass-period distribution for these stars. However, by assuming that the mass-period distribution is the same for M dwarfs as for more massive stars, we constrained the occurrence rate of planets relative to the FGK stars, taking into account possible differences in detectability between the two groups. Our results shows that the occurrence rate of gas giants within $2000$ days is $r=3$--$10$ times smaller for M dwarfs than FGK dwarfs (Fig.~\ref{fig:avni5}), with a two sigma limit $r>1.5$. A lower incidence of Jupiter mass planets around M dwarfs is predicted by core accretion models for planet formation (Laughlin, Bodenheimer, \& Adams 2004; Ida \& Lin 2005; Kornet \& Wolf 2006; Kennedy \& Kenyon 2008). Comparing with Figure 8 of Ida \& Lin (2005), we find that both the absolute and relative occurrence rates that we derive for Jupiter mass planets agree best with their standard model, in which disk mass is an increasing function of stellar mass ($\propto M_\star^2$). Kennedy \& Kenyon (2008) scale their disk mass $\propto M_\star$, but include a detailed calculation of the position of the snow line. Their Figure 7 shows that the probability of having at least one giant planet  is $6$ times lower for $0.4\ M_\odot$ star than a $1\ M_\odot$ star, within the range found in this paper. We can rule out a larger gas giant planet fraction for M dwarfs than for solar mass stars, as found by Kornet \& Wolf (2006) for models in which the disk parameters were independent of stellar mass. 

Our calculations can be improved in several respects. First, we neglected eccentricity when accounting for non-detections. This is reasonable for most values of $e$, since eccentricity has a large effect on detectability for $e\gtrsim 0.6$ (Endl et al.~2002; C04). However, the population of high eccentricity planets ($e>0.6$) is not well constrained. In addition, there are more subtle selection effects involving eccentricity. For example, Cumming (2004) showed that non-zero eccentricity enhances detectability for orbital periods longer than the time baseline of the data, introducing a bias in the longest period orbits towards systems with $e>0$. Further analysis is required to study the eccentricity distribution and the orbital period distribution of long period planets. Our data potentially allow us constrain the distribution of orbital periods beyond the 8 year time baseline of the observations, but this will require averaging over the range of possible eccentricities for those outer planets. We did not include multiple planet systems. This introduces some uncertainty in our derived distributions, since our technique identifies only a single planet. In a multiple system, this is the planet with the largest velocity amplitude, so that the distributions derived here are for the most detectable planet in a system. Finally, our method for including the upper limits involves dividing the data into either detections or non-detections, which depends on the choice of detection threshold. A better approach would be to calculate the probabilities directly and include them in the analysis (see eq.~[\ref{eq:app1}]).  Techniques to evaluate the relevant Bayesian integrals over the multidimensional parameter space have been discussed in the literature (Ford 2006; Ford \& Gregory 2007; Gregory 2007a, 2007b). This approach will allow orbital eccentricity, the uncertainty in parameters associated with long orbital periods, and multiple companions to be included.

\acknowledgements
We thank Peter Bodenheimer, Rychard Bouwens, Ting-Kuei Chou, Gil Holder, Greg Laughlin, Doug Lin, and Steve Thorsett for useful discussions. We gratefully acknowledge the dedication and support of the Keck Observatory staff. We thank NASA and the University of California for generous allocations of telescope time. The authors extend thanks to those of Hawaiian ancestry on whose sacred mountain of Manua Kea we are privileged to explore the universe with them. 
We are grateful for generous funding support through NASA grant NNG05GK92G to GWM, NSF grant AST-0307493 to SSV, NASA grant NNG05G164G to DAF, and the Cottrell Science Scholar program to DAF.
AC thanks Caltech Astronomy Department and the Kavli Institute for Theoretical Physics, Santa Barbara for hospitality while this work was in progress. Part of this work was completed while AC was supported by NASA at the University of California, Santa Cruz, through Hubble Fellowship grant HF-01138 awarded by the Space Telescope Science Institute, which is operated by the Association of Universities for Research in Astronomy, Inc., for NASA, under contract NAS 5-26555. AC is currently supported by an NSERC Discovery Grant, Le Fonds Qu\'eb\'ecois de la Recherche sur la Nature et les Technologies, and the Canadian Institute for Advanced Research, and is an Alfred P.~Sloan Research Fellow.

\appendix
\section{Correcting for the upper limits: A maximum likelihood approach}

\subsection{Likelihood function for detections and non-detections}

In this Appendix, we derive equation (\ref{eq:iterate}) for the completeness corrections starting with a maximum likelihood approach. We assume that the fraction of stars with a planet is $F_P$, with a mass-period distribution $f(M,P)$ normalized so that $\int d\ln M \int d\ln P\ f(M,P)=F_P$. The likelihood of the data for star $i$ is
\begin{equation}\label{eq:app1}
L_i=(1-F_P)q_i + \int d\ln M \int d\ln P\ f(M,P) p_i(M,P),
\end{equation}
where $p_i(M,P)$ is the probability of the data given a planet of mass $M$ and period $P$, and $q_i$ is the probability of the data given that no planet is present (see eqs.~[10] and [11] of C04). To determine $f(M,P)$, we maximize the total likelihood, which is the product over all stars of the individual likelihoods, $L=\prod_i L_i$. 

Note that equation (\ref{eq:app1}) for the likelihood assumes that each star has either no planet or one planet, i.e. that the presence of a planet with mass $M$ and period $P$ excludes the possibility of additional planets with other masses and periods. This is consistent with our search algorithm described in \S 2.3 which fits a single Keplerian orbit, and for multiple planet systems identifies only the planet with the largest radial velocity amplitude. The distribution $f(M,P)$ that we derive should therefore be considered as the mass-period distribution of the most detectable planet in a system. This is a close approximation to the true distribution since the number of multiple planet systems is $\approx 10$\% of the total (e.g. Marcy et al.~2005a). To include the possibility of multiple planets, the likelihood function can be derived by considering each bin in mass-period space as an independent Poisson trial (e.g., as Tabachnik \& Tremaine 2002), however this reduces to equation (\ref{eq:app1}) when the expected number of planets per star is $\ll 1$ (compare eq.~[10] of Tabachnik \& Tremaine 2002 with eq.~[\ref{eq:like}] below; see also Appendix B of Tokovinin et al.~2006 for a clear discussion).

In this paper, rather than evaluate $p_i$ and $q_i$ directly, we have classified each data set as either a detection or a non-detection. In the case of a detection, we make the approximation
\begin{equation}
q_i\approx 0,\hspace{1cm}p_i(M,P)\approx M_iP_i\delta(M-M_i)\delta(P-P_i),
\end{equation}
since we expect the likelihood to be strongly peaked near the best fit mass and period for a strong detection, with a vanishing probability that no planet is present. We write the mass, period, and velocity amplitude of detection $i$ as $M_i$, $P_i$, and $K_i$. For a non-detection, we expect
\begin{equation}
p_i(M,P)\propto\cases{{q_i} & $K<K_{{\rm up},i}$\cr
{0} & $K>K_{{\rm up},i}$\cr},
\end{equation}
since we are able to rule out velocity amplitudes above the upper limit $K_{{\rm up},i}$ but not below it, and it remains possible that the star has no planet.
Substituting these approximations into equation (\ref{eq:app1}) gives
\begin{equation}
L_i\propto \cases{{f(M_i,P_i)} & detection\cr {1-\int_{K>K_{{\rm up},i}}d\ln M\ d\ln P\ f(M,P)} & non-detection\cr}
\end{equation}
where we have used the normalization $\int d\ln M \int d\ln P\ f(M,P)=F_P$. The total likelihood is
\begin{equation}\label{eq:like}
L=\prod_{i=1}^{N_d}f(M_i,P_i)\ \prod_{j=1}^{N_\star-N_d}\left(1-\int_{K>K_{{\rm up},j}}d\ln M\ d\ln P \ f(M,P)\right),
\end{equation}
where $N_d$ is the number of detections out of $N_\star$ stars. Equation (\ref{eq:like}) is a two-dimensional generalization of the likelihood function of Avni et al.~(1980).

As a quick aside to gain some intuition, let's assume that $f(M,P)$ is a constant independent of $M$ and $P$. In addition, assume that $K_{\rm up}$ is the same for each star with a non-detection. Then,
\begin{equation}
L\propto F_P^{N_d}\ (1-kF_P)^{N_\star-N_d}
\end{equation}
where $k$ is the fraction of planets ruled out by the upper limit. To find the best fit value of $F_P$, we maximize $L$ by setting $dL/dF_P=0$. This gives
\begin{equation}\label{eq:intu}
F_P={N_d\over N_\star}k^{-1}.
\end{equation}
If the upper limit rules out the whole mass-period plane (in other words we can say that a star without a detection definitely does not have a planet) then $k=1$ and $F_P=N_d/N_\star$. However, if $k<1$ then we can exclude only a fraction $k$ of planets, and our estimate of $F_P$ must therefore be larger. For example, if a third of planets lie above the upper limit ($k=1/3$), we conclude that there are three times as many planets as we actually detect, $F_P=3N_d/N_\star$.

\subsection{Maximizing the likelihood: non-parametric approach}

To proceed further, one could divide the mass-period plane into a grid and solve for $f(M,P)$ in each grid cell (a non-parametric approach), or assume a parametric form for $f(M,P)$ and find the parameters that maximize the likelihood. We follow Avni et al.~(1980, \S Vb) which is to discretize $f$ at the locations of the detected planets, i.e.~write $\int d\ln M\ d\ln P\ f(M,P)=\sum f_i$, where we use the shorthand $f_i=f(M_i,P_i)$, the sum is over the detections, and the normalization is $\sum f_i=F_P$. This method gives $f(M,P)$ in a non-parametric way, but without binning the distribution. The total log likelihood is
\begin{equation}\label{eq:like2}
\log L=\sum_{i} \log f_i\ +\ \sum_{j} \log \left[ 1-\sum_{i, K_i>K_{{\rm up},j}} f_i\right],
\end{equation}
where the sums with index $i$ are over detections and the sum with index $j$ is over all non-detections.

We can now go ahead and maximize $L$ with respect to the $N_d$ values of $f_i$. We set $\partial L/\partial f_i=0$, which gives
\begin{equation}\label{eq:iterate1}
f_i=\left[\sum_{j,K_{{\rm up},j}<K_i}{1\over Z(K_{{\rm up},j})}\right]^{-1},
\end{equation}
where the sum with index $j$ is over all non-detections with upper limits that exclude a companion with the amplitude of detection $i$, and we define
\begin{equation}
Z(K_{\rm up})=1-\sum_{i,K_i>K_{\rm up}}f_i,
\end{equation}
which has a sum over all detections that have velocity amplitudes larger than the specified upper limit $K_{\rm up}$. 

Equation (\ref{eq:iterate1}) is an equation for $f_i$ which can be solved iteratively, as in \S \ref{sec:avni}. However, we proceed a little further in order to make connection with the heuristic derivation given in \S \ref{sec:avni}. We write equation (\ref{eq:iterate1}) as
\begin{equation}
1=f_i\sum_{j,K_{{\rm up},j}<K_i}{1\over Z(K_{{\rm up},j})},
\end{equation}
 and then sum both sides over the detections, from $i=1$ to $i=N_d$. After changing the order of the sums on the right hand side and simplifying, we find the result
\begin{equation}\label{eq:app2}
 N_\star=\sum_j{1\over Z(K_{{\rm up},j})}
\end{equation}
where the sum $j$ is over all non-detections. Using this to rewrite equation (\ref{eq:iterate1}), we find
\begin{equation}\label{eq:iterate2}
f_iN_\star=1+\sum_{j,K_{{\rm up},j}>K_i}{f_i\over Z(K_{{\rm up},j})},
\end{equation}
where the sum is over all non-detections $j$ which have upper limits that allow a companion with the same amplitude as detection $i$ to be present. Equation (\ref{eq:iterate2}) is therefore equivalent to equation (\ref{eq:iterate1}), and with the final definition $N_i=N_\star f_i$, reduces to equation (\ref{eq:iterate}) of \S \ref{sec:avni} when solved iteratively.

\subsection{The limit of small bin size}

One might worry that by following the distribution $f(M,P)$ only at the location of the detected planets, we are missing those areas of the mass-period plane in which there are no detections. In fact, we show now that the converged solution has non-zero values for $f$ only at the locations of the detected planets. We start with equation (\ref{eq:like}) and divide the mass-period plane into grid cells, labelled by $\alpha$ so that we will write $f(M,P)$ averaged over grid cell $\alpha$ as $f(\alpha)$. The grid cell containing detection $i$ we write as $\alpha_i$. The likelihood is then given by
\begin{equation}
\log L=\sum_{i} \log f(\alpha_i)\ +\ \sum_{j} \log \left[ 1-\sum_{\alpha, K(\alpha)>K_{{\rm up},j}} f(\alpha)\right].
\end{equation}
In the last term, $K(\alpha)$ is the velocity amplitude associated with grid cell $\alpha$, and so the sum is over only those grid cells that are constrained by the upper limit $j$. We assume that the grid cells are small enough that the entire grid cell is either excluded or allowed by the upper limit; in practice only a part of the grid cell may be excluded by the upper limit, but we ignore this here for clarity. Now, maximizing $\log L$ with respect to the set of $f(\alpha)$ by setting $\partial L/\partial f(\alpha)=0$, we find
\begin{equation}
f(\alpha)=N_d(\alpha)\ \left[\sum_{j, K_{{\rm up}, j}>K(\alpha)}{1\over Z(K_{{\rm up},j})}\right]^{-1},
\end{equation}
where $N_d(\alpha)$ is the number of detections in grid cell $\alpha$ and $Z(K_{\rm up})=1-\sum_{K(\alpha)>K_{\rm up}} f(\alpha)$. Following the same steps as in our derivation of equation (\ref{eq:iterate2}), we rewrite this as an iterative procedure, 
\begin{equation}
f^{n+1}={N_d(\alpha)\over N_\star}+\sum_{j,K_{{\rm up},j}>K(\alpha)}{f^n(\alpha)\over Z(K_{{\rm up},j})}{1\over N_\star},
\end{equation}
which should be compared with equation (\ref{eq:iterate2}). Now subtracting $f^n(\alpha)$ from both sides, we find
\begin{equation}
f^{n+1}(\alpha)-f^n(\alpha)={N_d(\alpha)\over N_\star}-f^n(\alpha)\sum_{j,K_{{\rm up},j}<K(\alpha)}{1\over N_\star Z(K_{{\rm up},j})},
\end{equation}
where we use a result analogous to equation (\ref{eq:app2}). We see that the converged solution ($f^{n+1}=f^n$) is
\begin{equation}
f(\alpha)=N_p(\alpha)\ \left[\sum_{j,K_{{\rm up},j}<K(\alpha)}{1\over N_\star Z(K_{{\rm up},j})}\right]^{-1}.
\end{equation}
Therefore, the converged solution has $f(\alpha)$ non-zero only in those grid cells which have detections ($N_d(\alpha)>0$). Taking bins small enough to contain either one or zero planets, we see that we need evaluate $f$ only at the locations of the detected planets, which is our starting point for equation (\ref{eq:like2}) (see also \S V.b of Avni et al.~1980, who show that a solution of eq.~[\ref{eq:like}] can be derived which is a sum over delta functions evaluated at the detected points).

\subsection{Likelihood function for power law mass and period distributions}

We now derive the likelihood $L$ for an assumed power law distribution $dN=f(M,P)d\ln Md\ln P=C\ M^{\alpha} P^\beta d\ln Md\ln P$, where $C$ is a normalization constant. This is used in \S \ref{sec:mp} to derive the best-fitting values of $\alpha$ and $\beta$ by maximizing $L$ with respect to these parameters. The calculation here is very similar to Tabachnik \& Tremaine (2002) except for our different form for $L$ (see discussion following eq.~[\ref{eq:app1}]). The log likelihood is
\begin{equation}\label{eq:like3}
\log L=\sum_i\left(\alpha\log M_i+\beta\log P_i+\log C\right)+\sum_j\log\left[1-C\ I_j\left(\alpha,\beta\right)\right]
\end{equation}
where 
\begin{equation}
I_j(\alpha,\beta)=\int_{K>K_{{\rm up}, j}} d\ln M\ d\ln P\ M^\alpha P^\beta.
\end{equation}
Setting $\partial L/\partial C=0$ gives the relation
\begin{equation}\label{eq:appC}
N_d=\sum_j {C\,I_j(\alpha,\beta)\over 1-C\,I_j(\alpha,\beta)}
\end{equation}
which can be solved to determine $C$ for given $\alpha$ and $\beta$. We calculate the integrals $I_j(\alpha,\beta)$ numerically, taking into account the variation of the upper limit $K_{\rm up}$ with orbital period. Following Tabachnik \& Tremaine (2002), after choosing the range of  $\ln P$ and $\ln M$ we are interested in, we refine it to just cover the range of periods and masses of the detected planets, as this maximizes the likelihood.

In \S \ref{sec:Mdwarfs}, we divide the stars into two groups based on their mass, and, assuming fixed values of $\alpha$ and $\beta$, fit separately for the two normalizations $C_1=f C$ and $C_2=C$, where $f$ is the relative planet fraction of group 1 (stellar masses $<0.5\ M_\odot$) compared with group 2 (stellar masses $>0.5\ M_\odot$). It is straightforward to show that in this case, $C$ and $f$ are determined by solving
\begin{equation}
N_1=\sum_{j(1)}{CfI_j\over 1-CfI_j};\hspace{1cm} N_2=\sum_{j(2)}{CI_j\over 1-CI_j}
\end{equation}
where $j(1)$ or $j(2)$ indicates a sum over the non-detections in group 1 or 2 respectively, and $N_1$ and $N_2$ are the number of detections in groups 1 and 2.

To get a feel for the solution, it is helpful to solve equation (\ref{eq:appC}) for the constant $C$ in the approximation that the integrals $I_j$ are the same constant for all stars with non-detections. This gives
\begin{equation}
f(M,P)d\ln M\ d\ln P=\left({N_d\over N_\star}\right){M^\alpha P^\beta\ d\ln M\ d\ln P\over \int_{K>K_{\rm up}}M^\alpha P^\beta\ d\ln M\ d\ln P}.
\end{equation}
The last term counts the number of constraining observations, analogous to the factor $k$ in equation (\ref{eq:intu}). In the case where we divide the stars into two groups, the  ratio of normalizations that maximizes the likelihood is
\begin{equation}
{C_1\over C_2}={N_1/N_{\star,1}\over N_2/N_{\star,2}}{I_2\over I_1}.
\end{equation}
The factor $I_2/I_1$ accounts for the relative detectability of planets in group 1 or 2. If detectability is the same for the two groups of stars, $I_1=I_2$, the estimate for $C_1/C_2$ corresponds to simply counting the relative fractions of detections.

\end{document}